\documentclass[prl,aps,showpacs,twocolumn]{revtex4}%
\usepackage{graphicx}
\usepackage{amsmath}
\usepackage{amsfonts}
\usepackage{amssymb}%
\providecommand{\U}[1]{\protect\rule{.1in}{.1in}}
\begin{document}
\title{Cooperative Lattice Dynamics and Anomalous Fluctuations of Microtubules}
\author{Herv\'{e} Mohrbach$^{1,2}$, Albert Johner$^{2}$ and Igor M. Kuli\'{c}$^{2}$}
\email{[Email: ]kulic@unistra.fr}
\date{\today}

\begin{abstract}
Microtubules have been in biophysical focus for several decades. Yet the
confusing and mutually contradicting results regarding their elasticity and
fluctuations have shed some doubts on their present understanding. In this
paper we expose the empirical evidence for the existence of discrete
GDP-tubulin fluctuations between a curved and a straight configuration at room
temperature as well as for conformational tubulin cooperativity. Guided by a
number of experimental findings, we build the case for a novel microtubule
model, with the principal result that microtubules can spontaneously form
micron size cooperative helical states with unique elastic and dynamic
features. The polymorphic dynamics of the microtubule lattice resulting from
the tubulin bistability quantitatively explains several experimental puzzles
including anomalous scaling of dynamic fluctuations of grafted microtubules,
their apparent length-stiffness relation and their remarkably curved-helical
appearance in general. We point out that tubulin dimers's multistability and
its cooperative switching could participate in important cellular processes,
and could in particular lead to efficient mechanochemical signalling along
single microtubules.

\end{abstract}
\affiliation{$^{1}$Groupe BioPhysStat, Universit\'{e} Paul Verlaine-Metz, 57078 Metz,
France }
\affiliation{$^{2}$ CNRS, Institut Charles Sadron, 23 rue du Loess BP 84047, 67034
Strasbourg, France }

\pacs{87.16.Ka, 82.35.Pq, 87.15.-v}
\maketitle

\section{Introduction}

Microtubules (MTs) are fascinating biological macromolecules that are
essential for intracellular trafficking, cell division and maintaining the
cell shape. They display unique elastic and dynamic behavior, inherited by
their complex self-assembling nanotube structure. Surprisingly, despite an
enormous accumulation of knowledge about their structure, the MTs static and
dynamic properties have challenged all attempts of a fully coherent
interpretation. In this paper we develop the line of thoughts leading towards
a new theory that provides a more comprehensive understanding of these MT's
properties. The fundamental result is that stabilized MTs spontaneously form
large scale superhelices of micron size pitches and diameters. The MT's
super-helical structure turns out to be a consequence of a cooperative
interaction between its individual subunits that can sustain several stable
curved conformations. Cooperativity of fluctuating internal degrees of freedom
in combination with the cylindrical MT symmetry lead to a helical state with
very unique characteristics in the world of macromolecules: MTs are helices
that are permanently but coherently reshaping -i.e. changing their reference
ground state configuration- by thermal fluctuations. In particular when
clamped by one end MTs undergo an unexpected zero energy motion. As we will
see, this could be the key for a consistent interpretation of certain
challenging experimental results not captured by the conventional scenarios
and models. It is worth remarking, that the large majority of experiments
probing the above mentioned properties are in-vitro experiments on MTs with
stabilizing drugs. We stress this point, as the experimentally prevalent
presence of stabilizing agents is not innocent and could modify MT's
properties. Nevertheless there are reasons to believe that the theory
developed here might be valid also for non treated MTs.

Before embarking on the road to a "polymorphic MT theory", we first review the
conventional understanding of the common MT properties as well as some key
experiments which will be our necessary guides towards the model we propose.
This paper which extends and deepens a previous short presentation of the idea
of polymorphic MTs \cite{PAPER1} is written in two parts. In the first part,
we provide conceptual and graphical explanations of the ideas behind the model
and investigate consequences of the here developed polymorphic MT model. We
hope that this part is self consistent and didactic enough, so that a general
reader can grasp the basic underlying ideas. In the second more technical
part, the mathematical model is developed and quantitative results are
presented. The details and derivations are left to an extensive appendix.

\section{Short review of what we understand about microtubules}

Microtubules are cytoskeletal protein filaments of eukaryotic cells fulfilling
different structural and mechanical functions in the cell: MTs act as
"cellular bones" strongly influencing the cell shape, constitute the main
routes for molecular motor mediated intracellular cargo transport \cite{Genref
MTs, GeneralAmos} and perform other important tasks like stirring the
cytoplasm \cite{MTStirringRod}. Besides, MTs play a central role in the
assembly of the mitotic spindle during cell division and are at the heart of
the functioning of cilia and flagella \cite{Alberts}. This versatility of MTs
in a variety of biological functions mainly relies on their unique high
stiffness and on their dynamics of assembly and disassembly. The high rigidity
of MTs (similar to hard plastic) is due to their structure that is known in
exquisite detail from $3$D electron microscopy reconstructions
\cite{NogalesMTubule,NogalesMT2} : MTs are hollow tubes whose walls are formed
by assembly of a variable number $N$ of parallel protofilaments (PFs). The PFs
themselves are built by head-to-tail self association of the $\alpha\beta
$-tubulin heterodimer protein subunit (yielding a polarity to MTs) whose
structure has been resolved by electron crystallography
\cite{NogalesTubulin,NogalesTubulin2}.

In vivo, MTs most commonly appear with $13$ PFs \cite{Bouchet} (although there
are exceptions depending of the cell type), whereas in vitro a variety of
structures with PF number ranging from $N=9$ to $16$ were observed
\cite{Wade,Chretien}. Transitions of different lattice types (mostly with the
gain or loss of one PF) within a single MT are also frequent
\cite{Ray,ChretienMetoz,ChretienFuller}. The MT lattice can accommodate for
these different structures by twisting the PFs around the central MT axis
although this process is energetically costly. The typical lattice twist
repeat lengths (pitches) are $:+3.4,+25,-6\mu m$ for $N=12$, $13$ and $14$ PF
MTs respectively, with +/- denoting right/lefthanded twist \cite{Wade,
Chretien, Ray, ChretienMetoz, ChretienFuller}. As we will see the lattice
twist is an essential property of the model that we will propose later on. Any
deviation from the most frequent, energetically favorable and thus less
twisted configuration $N=13$ implies an internal prestrain in the MT lattice.
The latter stress can locally deform the end portions of the lattice or even
destabilize it \cite{HunyadiMechanical}.

Another internal prestrain in the MT lattice is believed to be caused by the
tubulin subunit which upon incorporation into the lattice hydrolyzes a bound
GTP\ molecule converting it quickly into a GDP-tubulin form which has the
tendency to form a curved state - curling radially away from the axis
\cite{Mandelkow,NogalesRings}. The constraint imposed by the MT lattice
however maintains the GDP-tubulin dimers in a straight unfavorable state - in
turn trapping mechanical prestrain. In this conventional view MTs are seen as
internally prestrained but intrinsically straight Euler beams. The GDP-tubulin
prestrain is also believed to trigger rapid depolymerization called the
polymerization "catastrophe" \cite{Mitchison}. The stability of MT is
regulated either by the presence of a thin layer of yet unhydrolyzed GTP
tubulin dimers at the growing MT end (the so called GTP-cap-model,
\cite{EricksonCap,JanosiCap}) or by the binding of MT-associated proteins
(MAP) or of drugs such as taxol.

\section{What we don't understand about microtubules}

\subsection{Mechanical properties of stabilized MTs}

To investigate the mechanical properties of the MT lattice the presence of the
polymerization dynamics is often an obstacle. As previously mentioned it can
be switched off by stabilizing agents like taxol or MAPs. In the bulk of the
available in vitro studies taxol stabilization has been the experimental
method of choice for investigating the elastic properties of MT. It is
believed that taxol maintains tubulin dimers in an approximately straight
conformation and thus prevents depolymerization \cite{ArnalTaxol, AmosTaxol,
Xiao}. However, direct EM investigations of single taxol-GDP-PFs
\cite{Multistable Tub EM} reveals a more complex and interesting picture. A
taxol stabilized single PF can in fact coexist in several conformational
states with comparable free energies: a straight state $\kappa_{PF}%
^{st}\approx0$ and a weakly curved state with intrinsic curvature $\kappa
_{PF}^{wc}\approx1/250nm$ (see Fig. 1a). A third highly curved state with
$\kappa_{PF}^{hc}\approx1/20nm$ additionally appears after longer observation
periods. Notably Elie-Caille et al. \cite{Multistable Tub EM} pointed out a
cooperative nature of the straight to curved transition within single PFs.
These important findings -several conformational tubulin dimer states and
cooperative interaction between them- will be one central ingredient of our
model later on.

Going from a single protofilament to the whole tube, a central mechanical
property of interest often measured for stabilized MTs, is its persistence
length defined as $l_{p}=B/k_{B}T.$ Here $B\propto Y$ stands for the flexural
rigidity which is for an isotropic beam proportional to its elastic\ Young
modulus $Y$. The persistence length is the length scale characterizing the
filament's resistance to thermally induced bending moments. Several
experimental approaches have been developed to measure bending stiffness of
taxol stabilized MTs. One method consists in measuring MT's thermal shape
fluctuations via dark-field microscopy \cite{GITTES,Venier} or fluorescence
light microscopy \cite{Mickey}. Alternatively in several other experiments,
$l_{p}$ has been determined by applying controlled bending forces via
hydrodynamics \cite{Venier}, optical tweezers
\cite{Felgner,Kukimoto,ActtiveMTBending, TAKASONEbuckling} and atomic force
microscope tips \cite{Kis}. Interestingly the authors in \cite{GITTES}
observed that stabilized MTs are not perfectly straight, but contain some form
of quenched curvature disorder which needs to be subtracted from the
measurements. In the same vein, the observations from dark-field images of the
thermal fluctuations of the free end of axoneme-bound MTs, show that taxol
stabilized MTs adopt a three-dimensional helicoid structure \cite{Venier} (we
will return to this point later on).

The persistence length obtained from these different experiments is set in the
range of $1$ to $8$ $mm$ with some of newer studies going down to $0.1mm$ (cf
below). For a standard semiflexible polymer (like e.g. actin or DNA) we expect
$l_{p}$ to be a material constant, in particular independent of the filament's
length. However, for MTs, the experimental $l_{p}$ data are not only highly
scattered but extremely confusing on this point. That the persistence length
could indeed depend on the MT length was first mentioned and observed in
references \cite{ActtiveMTBending, TAKASONEbuckling} and confirmed by other
experimental measurements - probing either the thermal movement
\cite{GlidingAssay} or the active bending deformations by electrical fields of
individual taxol stabilized MTs gliding on a kinesin-coated surface
\cite{GlidingAssay2,GlidingAssay3}. In particular these techniques gave for
short MT segments with submicron length a persistence length between $0.08$
and $0.24$ $mm$.

This intriguing "length dependent stiffness" was also investigated by
Pampaloni et al who measured the lateral fluctuations of MTs grafted to a
substrate \cite{Pampaloni}. These authors found a $l_{p}$ falling within a
range of $0.11$ to $5.04$ $mm$ for MT lengths varying from $2.6$ to $47.5$
$\mu m$ - with a strong linear length-$l_{p}$ correlation. A similar
experiment done in \cite{Taute} measured the longest relaxation time
$\tau_{\max}\left(  L\right)  $ of MTs for various MT lengths $L$. It was
found that MTs exhibit unusually slow thermal dynamics compatible with
$\tau_{\max}\propto L^{3}$ (cf. Fig 9) - in sharp contrast to standard
semiflexible filaments with $\tau_{\max}\propto L^{4}$. In \cite{Brangwyne},
the relaxation time for long MTs ($L>10\mu m$) extracted from a 2-D shape
analysis of taxol stabilized fluorescent MTs shows an anomalous dynamics on
short scales as well. Furthermore it has also been experimentally found that
the rigidity of MTs depends on their growth velocity \cite{Janson}. All these
experiments measuring the bending stiffness and the bending dynamics brought
the community to the conclusion that the "beam of Life" cannot be seen as a
simple Euler beam. Its complex internal structure should determine its elastic
and dynamical properties. But which internal mode contributes to the now
obvious elastic complexity, is the important question - a convincing answer of
which is still missing.

\begin{figure}[ptbh]
\begin{center}
\includegraphics[
width=3.5 in ]{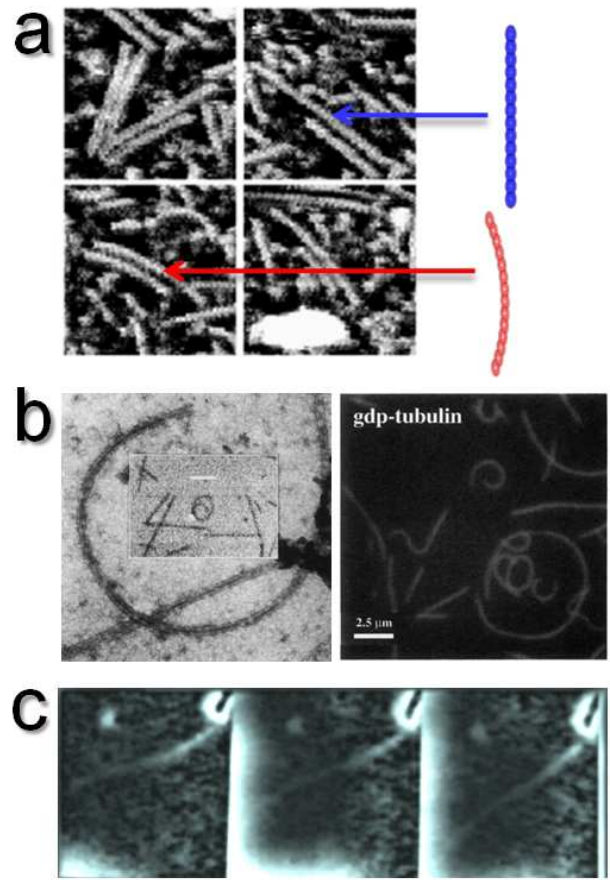}
\end{center}
\caption{The empirical evidence for tubulin bistability: (a) A single taxol
stabilized protofilament can coexist in a straight $\kappa_{PF}\approx0$ and a
slightly curved $\kappa_{PF}\approx1/250nm$ state (reproduction from
\cite{Multistable Tub EM}) (b) Taxol stabilized microtubuls in gliding assay
experiments can switch to a stable circular state and move on circular tracks
(from \cite{Amos}\cite{ValeCoppin}). Microtubules are occasionally observed to
switch back and forth between the circular and straight states. (c) Taxol
stabilized microtubules form a three-dimensional helicoid structure with a
$15\mu m$ pitch (from \cite{Venier}).}%
\end{figure}

\subsection{Helices and Rings}

Even more intriguing than the issue of MT elasticity is perhaps the question
:\ What is the ground state of a MT? While the naive answer - a straight rod -
would be the most accepted view, a number of experiments put this mundane
simplistic picture in doubt. For instance, wavy sinusoidal and circular shapes
are frequently observed when MTs are adsorbed to glass surfaces or confined
between them. In this confined case the Fourier mode analysis of MT
deformations systematically reveals that a few discrete modes have a larger
amplitude than the fluctuations around them \cite{GITTES,Brangwyne,Janson},
cf. also \cite{VanHeuvel} -supplementary material. This is a strong hint
towards the presence of some type of "frozen in" curvature - dynamically
quenched on experimental timescales. Likewise in motility assays, it is often
seen that when MTs glide over a surface coated with molecular motors they
follow wavy sinusoidal and often circular tracks \cite{Amos,ValeCoppin}, cf.
Figs. 1b. In this context, particularly interesting is the observation by Amos
\& Amos \cite{Amos} of the formation of permanently circling MTs (see Figs
1b). These unusual stable circular gliding structures persisted for many
cycles and occasionally straightened again. As written by the authors
\cite{Amos}, this suggests that "an intact tubular polymer is capable of
holding more than one conformational state without the help of an external
force". From EM images it was inferred that the underlying mechanism stems
from the balance of individual taxol-GDP-tubulin dimers between two or more
different stable conformations \cite{Amos EM}.

An even more clear hint towards the real nature of the "frozen in "\ curvature
was -as already mentioned- discovered by Venier et al \cite{Venier} (Fig.1c)
who described stabilized MTs as \textit{wavy periodic shapes} with a half
period of about $7-8\mu m$. This observation combined with the fact that MTs
often went out of focus, lead the authors to "suggest that taxol-treated
microtubules may adopt a three-dimensional helicoid structure of $15\mu m$ pitch".

Therefore it seems that the MT curvature is a persistent attribute attached to
its lattice. Any serious attempt to fully understand the complexity of
MT\ elasticity can not avoid the question of its ground state. Indeed
understanding fluctuations around a particular state will be futile\ as long
as the origin of the state itself remains in the dark.

\subsection{The Soft Shear Model}

A first theoretical attempt to cope with some aspects of the described MT
mechanical complexity was the "soft shear model" (also called "Timoshenko beam
model", or "anisotropic composite material model")
\cite{Kis,Pampaloni,Frey,KulicMTshear}. In this model the MT\ is considered as
an anisotropic fiber-reinforced material \cite{Kis,Pampaloni} with the tubulin
protofilaments acting as strong fibers weakly linked with easily shearable
inter-protofilament bonds. Some specific equilibrium statistical and
mechanical properties of that model were investigated in
\cite{Frey,KulicMTshear}. An interesting peculiarity and inherent consequence
of this model is that any local lattice deformation gives rise to a long
distance curvature relaxation \cite{KulicMTshear} and can lead to a long range
interaction along the MT contour. This aspect of the "soft shear model" (SSM)
is in phenomenological agreement with cooperative deformations induced by
enzymes like katanin. Furthermore this model predicts a length-dependent
persistence length which approximately resembles the measured behavior
\cite{Pampaloni,Taute}. However in detail it suffers a number of difficulties
and inconsistencies in particular :

-The ground state of SSM is straight - in conflict with the helical ground
state observation \cite{Venier}.

- The SSM\ does not allow for lattice multistability as observed by Amos\&
Amos \cite{Amos}.

- The predicted value of the shear modulus is extremely small
\cite{Pampaloni,Taute,KulicMTshear} ($10^{5}-10^{6}$ times smaller than the
Young modulus). This would imply very strong shearing in bent microtubule
structures. This however is unsupported by other experimental evidence. Indeed
observations of straight and highly bent MTs shows that bending does not
significantly modify the relative position of the inter-protofilament bonds
\cite{Chretienlateralbonds}.

-The dynamics of clamped MTs \cite{Pampaloni,Taute} does not come naturally
out from the SSM. To reach agreement and fit the experimental dynamics the
shear model needs to introduce an ad hoc internal dissipation of unclear origin.

- For short MTs ($<4\mu m$) that model is very far off and disagrees with the
"plateau" region of the bending stiffness vs length relations ( cf. Fig 3 in
\cite{Taute}).

A careful reanalysis of clamped MT experiments, cf. Figs. 2,3 in
\cite{Pampaloni,Taute}, reveals two features not captured by the SSM : the
persistence length scales for large $L\ $approximately as $\sim L$ (without
signs of saturation) while the relaxation times scale as $\sim L^{3}$. This
exotic behavior naively suggests the presence of a limited angular hinge at
the MT\ clamping point. On the other hand artifacts that could trivially lead
to a \textquotedblright hinged behavior\textquotedblright\ (like loose
MT\ attachment and punctual MT\ damage) were specifically excluded in
experiments \cite{Pampaloni,Taute}.

Facing all these obstacles it becomes increasingly clear that the solution to
all puzzles requires a cut and a radically different hypothesis.

\section{Idea of Polymorphic Microtubule Dynamics}

The new scenario proposed here is based on the hypothesis of
\textit{cooperative} \textit{internal} MT lattice dynamics. The two central
assumptions of our model are as follows:

(I) The taxol-GDP-tubulin dimer is a conformationally multistable entity and
fluctuates between at least 2 states on experimental time scales :\ a straight
and an outwards curved state (Figs. 1a and 2a).

(II) There is a nearest-neighbor cooperative interaction of tubulin states
along the PF axis.

Note that assumption I is very different from the conventional picture where
GDP-tubulin has only \textit{one} energetically favorable (curved) state. We
will show that a model based on I and II straightforwardly leads us to the
very origin of MT (super) helicity and provides a coherent explanation for
static and dynamic measurements in thermal fluctuation experiments.

In contrast to the soft shear model, the present model is elastically
isotropic but the monomer curvature is bistable. As we will see, the ground
state in this new model is a highly degenerate 3 dimensional helix fluctuating
between many equivalent configurations.

\begin{figure}[ptbh]
\begin{center}
\includegraphics[
height=3.979in,
width=3.1081in
]{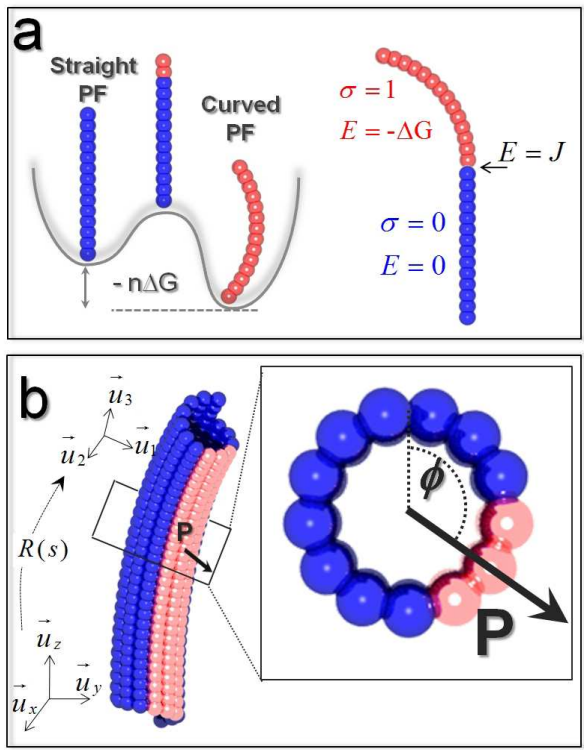}
\end{center}
\caption{Elements of the "polymorphic tube model". (a)\ The GDP-tubulin
protofilaments can fluctuate cooperatively between two discrete states. The
curved $\sigma=1$ state is energetically preferred over the straight state
$\sigma=0$ with an energy gain $E=-\Delta G$. The junction between straight
and curved states along the same protofilament are penalized by a coupling
constant $E=+J.$ b) Competition between tubulin switching energy and elastic
lattice strain energy leads to spontaneous symmetry breaking: MT bends to a
randomly chosen direction and assumes a non-zero polymorphic order parameter
$P$. The energy becomes invariant with respect to an arbitrary rotation of the
polymorphic phase angle $\phi$.}%
\end{figure}

\subsection{Conformational Symmetry Breaking and Helix Formation}

In this chapter we will give a simple pictorial panorama over the consequences
of assumptions I and II. What happens when tubulin dimers obeying assumption
I\ are trapped in the circularly symmetric MT lattice? Starting from
assumption I we imagine that the straight and curved GDP-tubulin state have a
certain energy difference $\Delta G>0,$ with the curved state being slightly
more favorable. The strict preference for the curved state is however only
true if the tubulin dimers are free i.e. not confined to the lattice. The
situation becomes more interesting once they are incorporated into
the\ lattice. Obviously the outwards bending tendency of the curved state is
in conflict with the geometry of the lattice. Switching a dimer on one
MT\ side will frustrate the dimers on the opposite side of the tube and
prevent them from switching at the same time. On the other hand the direct
neighbors of the curved dimer (on the same MT side) will profit and switch
easier to the curved state, as the lattice is already slightly "pre-bent" in
the correct direction. This peculiar interplay of negative and positive
interaction gives rise to a clustering of curved dimer states into a single
block on one side of the tube cf. Fig. 2b.

This immediately raises the question about the orientation of this curved
dimer block. Which MT\ side will be selected, and in which direction will the
MT\ overall bend, can only be decided by the process of \textit{spontaneous
symmetry breaking}. That is, if the ground state is a curved dimer block, it
will be a highly degenerate state (see Fig. 3). In turn it can be expected to
thermally move through the lattice at next to no energy cost (apart from some
friction)! This is the most essential feature of the present model.

So far we have considered only a single MT\ cross-section. If the assumption
II (cooperativity) would not hold the curved state blocks would pick their
sides at each section completely independently. The tube would locally curve
in random uncorrelated directions and the effect of tubulin multistability
would stay essentially invisible on the larger scale. However according to
assumption II\ the blocks become correlated in orientation and prefer to stack
on the top of each other. This then leads to macroscopically a bent - in fact
a circular MT - if the PF\ were not twisted around the central MT\ axis (see
Fig. 2b).\ This provision brings us to the last interesting point. As already
mentioned MT lattices are generically twisted i.e. their PFs are not strictly
parallel. With a cooperative interaction along the PFs - which are now
twisting around the tube axis - the final product of assumption I\ and II will
be a long pitched helix! The pitch of the helix should coincide with the
lattice twist repeat length $:+3.4,+25,-6\mu m$ for $N=12$, $13$ and $14$ PF
MTs respectively.\ The created "polymorphic helix", however will not be unique
and will be able to reshape between its $N\ $indistinguishable orientation states.

\begin{figure}[ptbh]
\begin{center}
\includegraphics[
height=3.6331in,
width=3.1081in
]{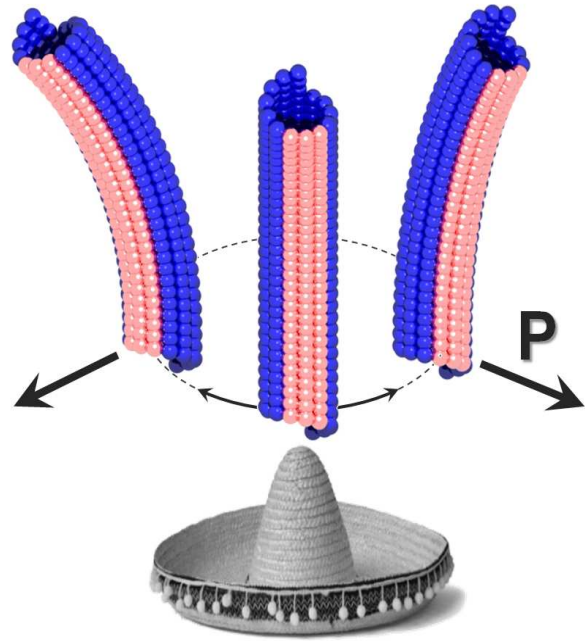}
\end{center}
\caption{The straight state of the microtubule becomes unstable and forms a
spontaneous bend with fixed curvature pointing towards a randomly chosen
direction. The microtubule can assume one of the N degenerate ground states
and switches between them at no energy cost - the effective potential has a
shape reminiscent of a Mexican hat.}%
\end{figure}

When we graft one end of such a polymorphic helix onto a surface (as e.g.
performed in \cite{Pampaloni,Taute}) the helix will still be able to switch
between the equivalent orientations and perform a motion that we call
"wobbling" (see Fig. 4). It is exactly this type of motion that can give rise
to the static and dynamic effects measured in \cite{Pampaloni,Taute}.

To see this we can approximate the movement of a clamped polymorphic helix
that is switching between its equivalent ground states with a "rigid conical
rotor" (see Fig. 4). For such a rotor the transverse displacement $\rho$ of
the MT end grows linearly with its length $L.$ Using the naive definition of
persistence length $l_{p}\approx L^{3}/3\left\langle \rho^{2}\right\rangle $
(where $\left\langle {}\right\rangle $ is the ensemble average) and the fact
that $\left\langle \rho^{2}\right\rangle \propto L^{2}$ this apparent
persistence length becomes proportional to the length $l_{p}\propto L.$ This
scaling (cf. Fig. 11) is in agreement with the experimental results in
\cite{Pampaloni,Taute} giving us the first hint that the model is on the right
track. Appendix E comments on the robustness of this conical hinge-like motion
against a limited local hindrance of the wobbling mode due to the adsorbed
part of the grafted MT.

A further encouragement comes from the study of the dynamics of the model.
Making again the approximation of a rigid conical rotation (induced by
wobbling) the observed unusual scaling of the longest relaxation time
$\tau_{\max}\propto L^{3}$ can also be easily understood. In fact a slender
object of length $L$ rotated along a conical surface has a friction constant
$\xi_{rot}\propto L^{3}$. In turn the longest relaxation time is given by
diffusional equilibration time on the cone, i.e. $\tau_{\max}\propto\xi
_{rot}/k_{B}T\propto L^{3}$ (cf. Fig. 12). It turns out that the model
correctly predicts both the exponent and the prefactor of the experimental
relaxation time. \begin{figure}[ptbh]
\begin{center}
\includegraphics[
width=3.0in ]{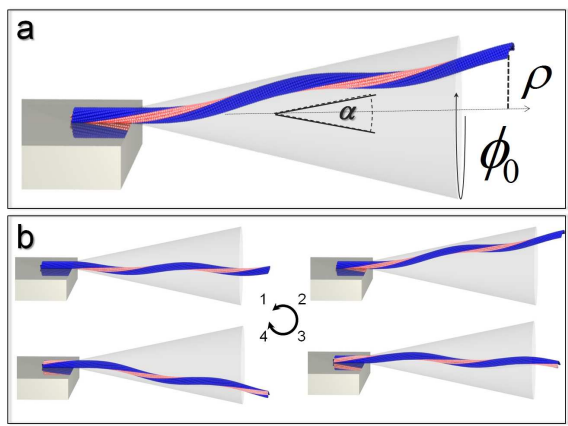}
\end{center}
\caption{Clamped polymorphic microtubule with intrinsic lattice twist attached
at its end to a substrate \cite{Pampaloni}\cite{Taute} performs a peculiar
movement. It forms a polymorphic helix with N degenerate ground states and
switches between them at no energy cost. The approximately conical motion with
opening angle $\alpha$ leads to anomalous lateral fluctuations $\left\langle
\rho^{2}\right\rangle \sim\alpha^{2}L^{2}$ - radically different from all
other semiflexible filaments. }%
\end{figure}

In addition it is further reassuring that the helical ground state(s) of the
model provides a rationale for the observation of MT\ helices by Venier et al.
\cite{Venier}.

\subsection{In Vivo Implications of Microtubule Polymorphism}

The bulk of observations cited here is made in vitro on taxol stabilized MTs.
It is known that taxol inhibits MT disassembly by maintaining the tubulin
dimer in a state that strongly favors polymerization. But from the experiments
reported above it is also clear that taxol does not suppress all tubulin
conformational changes : taxol-GDP-tubulin dimer has multiple stable
conformations \cite{Multistable Tub EM}. How do these in vitro findings relate
to MTs in vivo?

It is a common empirical observation in many in vivo systems that despite
their high bending stiffness MTs are often seen curved or highly wavy on
micron scales
\cite{Brangwyne,BicekInvivo,BrangwynneWavyBucking,Borisy,Kaech,SamsonovTau}.
For instance in \cite{BicekInvivo}, highly sinusoidal MTs on the periphery of
living LLC-PK1 cells were observed. These shapes are usually explained as a
consequence of motor induced buckling opposed by a gel-like environment that
leads to finite wavelength- buckling \cite{BrangwynneWavyBucking}. While the
"buckling in a gel" interpretation is physically appealing a closer look at
the data in \cite{BrangwynneWavyBucking} (in particular the accompanying movie
material) reveals an absence of correlation in buckling \textit{directions} of
neighboring MTs. This observation puts a question mark on the participation of
a background continuum gel - as in this case the strains in the gel would
necessarily propagate to neighboring microtubules and lead to spatially
correlated buckling events - in sharp contrast with observations. \ Therefore
we are left with a robust phenomenon of sinusoid, constant wavelength MTs -
however without a definite interpretation so far.

The phenomenology of stable rings \cite{Amos} and wavy sinusoid MTs forming in
gliding assays \cite{ValeCoppin} is strikingly similar and visually
indistinguishable from the pure in vivo observations
\cite{Brangwyne,BicekInvivo,BrangwynneWavyBucking,Borisy,Kaech,SamsonovTau}.
Indeed, in both situations highly curved lattice states of very similar
magnitudes $\kappa_{MT}\approx1-2\mu m^{-1}$ are readily observed. This
analogy between in vivo and vitro cases suggests that tubulin dimers - both in
vivo and taxol stabilized (in vitro) posses an identical highly curved state
($\kappa_{PF}^{h.c.}\approx1/20nm$). This highly curved state appears to be
activated within the lattice only under compressive loads and seems to be a
universal property of GDP tubulin itself - independent of taxol stabilization.

On the other hand the weakly curved state of taxol-GDP-dimer ($\kappa
_{PF}^{w.c.}\approx1/300nm$)) is soft enough to be activated by thermal
fluctuations. The empirical evidence for the weakly curved state in vivo is
far less obvious than for the highly curved state. The slight deformations
induced by the former would be more difficult to distinguish from other MT
deforming effects in living cells like motors, polymerization forces, presence
of lattice defects, bundling and microtubule associated protein action.
However the observed phenomenology of length dependent persistence length of
MTs growing from centrosomes in egg extracts \cite{Keller} (no taxol present)
resembles qualitatively and quantitatively the in vitro observations
\cite{Pampaloni,Taute}. This is one more indication that the dynamic
MT\ polymorphism could persist also in vivo.

At least two possible biological implications of the helicoidal polymorphic MT
nature come straight to mind. First, a curved or helical beam under
compressive load responds like a mechanical spring and is therefore much
softer (in tension, compression and bending) than a straight beam. Therefore a
network of helical MTs might be important for the overall mechanical
compliance of the cell. A perfectly straight MT buckled by an extracellular
load would be much more susceptible to mechanical failure and depolymerization
than a soft compliant helix. Second, helical shapes are geometrically
(topologically) prevented from side-by-side aggregation, can thus evade the
formation of bundles and instead form lose networks with much fewer contacts.
It appears that a tuned helicity (that can be switched on or off) could be a
good mechanical control parameter for the formation of different cytoskeletal
structures. While the bulk of the cytoplasmic MTs is preferably in the lose
network state (favored by helicity) in occasional situations like the neuronal
axons straight aligned MTs are required. In such a situation the bundling
could be triggered by switching the lattice to the straight state. Remarkably
in the process of axonal retraction the straight axonal bundle is destabilized
(and eventually contracted towards the cell soma)\ by an apparent transition
of the MTs to a wavy coiled state \cite{Baas} very reminiscent in shape to
single wavy MTs from cell soma
\cite{Brangwyne,BicekInvivo,BrangwynneWavyBucking,Borisy,Kaech,SamsonovTau}.

In general MT's polymorphism might have other, less obvious biological
implications that still have to be identified. In particular a more
speculative possibility is that the tubulin's allosteric multistability might
also be a piece in the puzzle of MT "catastrophes". A cooperative curvature
switch might trigger a transition from growth to depolymerization. Maybe the
most fascinating aspect could lie in the possibility of signal transmission
along single MTs via a long range conformational switch. If our model is
correct this is the most inherent and distinct consequence of the underlying mechanism.

We conclude here the qualitative description of essential ideas behind the
polymorphic tube model. In the following we switch gears and present in more
quantitative detail the mathematical model. The mathematically less inclined
reader is invited to fly over some figures, comparisons with experiments,
maybe pick up additional concepts (such as polymorphic defects and their
dynamics) and jump to the perspectives section which will underline the
essential biological consequences.

\section{The Polymorphic Tube Model}

Until now the discussion was qualitative and in the following we build the
mathematical model of the polymorphic MT. We will provide quantitative
arguments and model experimental data, justifying thus a posteriori the
previous discussion. In this section, we will focus on the thermally induced
weakly curved state in taxol-stabilized MTs and leave the case of mechanically
induced highly curved MTs for further works.

We model the GDP-tubulin dimer state by a two state variable $\sigma
_{n}\left(  s\right)  =0,1$ which corresponds to the straight and$\ $curved
state at each lattice site. The $n=1,...N$ \ is the circumferential PF index
and $s\in\left[  0,L\right]  $ is the longitudinal position variable along the
MT centerline. We recall that our model is based on the following assumptions:

(I) The taxol-GDP-tubulin dimer fluctuates between 2 states - straight and
curved - (Fig. 2a) with an energy difference $\Delta G>0$ favoring the curved
state. The energy density resulting from the switching of tubulin dimers (at a
given MT section) is then given by
\begin{equation}
e_{trans}\left(  s\right)  =-\frac{\Delta G}{b}\sum\nolimits_{n=1}^{N}%
\sigma_{n}(s) \label{e_trans}%
\end{equation}
with$\ b\approx8nm$ the dimer length.

(II) There is an Ising type nearest-neighbor cooperative interaction of
tubulin states \textit{along} the PF axis with an interaction energy $J>0$
favoring nearest neighbor dimers on the same PF to be in the same state. This
leads to the interaction energy density:%

\begin{equation}
e_{inter}\left(  s\right)  =-\frac{J}{b}\sum\nolimits_{n=1}^{N}\left(
2\sigma_{n}\left(  s\right)  -1\right)  \left(  2\sigma_{n}\left(  s+b\right)
-1\right)  \label{eJ}%
\end{equation}
The last term missing in our description is the elastic energy density of the
MT lattice. For a usual isotropic Euler beam the material deformations
$\varepsilon$ are related to the centerline curvature vector $\vec{\kappa}$
via $\varepsilon=-\vec{\kappa}\cdot\vec{r}$ with $\vec{r}$ the radial material
vector in the cross-section. For a polymorphic MT, modelled as a continuum
material made of $N$ PFs ($N=11-16$), its actual deformation will depend on
the polymorphism-induced prestrain $\varepsilon_{pol}.$ In this case the
elastic energy density of the MT can be written as
\begin{equation}
e_{el}\left(  s\right)  =\frac{Y}{2}\int_{R_{i}}^{R_{o}}\int_{0}^{2\pi}\left(
\varepsilon-\varepsilon_{pol}\right)  ^{2}rdrd\alpha\label{e_el}%
\end{equation}
where the integration in $e_{el}$ goes over the annular MT cross-section with
$R_{i}\approx7.5nm,$ $R_{o}\approx11.5nm$ the inner and outer MT radii
respectively. The prestrain $\varepsilon_{pol}$ is a function of the
polymorphic state of the tubulin dimers. Its definition requires a
decomposition of the tubulin dimer in an inner part (facing the tube axis) and
an outer part (facing from the tube axis outwards), cf. Fig 5. We assume that
each curved dimer state generates a positive prestrain $+\varepsilon_{PF}$ on
its inner part and an equal but negative prestrain $-\varepsilon_{PF}$ on its
outer part. We can then write $\varepsilon_{pol}\left(  s,r,\alpha\right)
=\varepsilon_{PF}\sigma_{n}\left(  s\right)  $ $\left[  I_{[R_{i},R_{o}%
-d_{PF}/2]}\left(  r\right)  -I_{[R_{o}-d_{PF}/2,R_{o}]}\left(  r\right)
\right]  $ $\cdot I_{[\frac{2\pi}{N}n+q_{0}s,\frac{2\pi}{N}\left(  n+1\right)
+q_{0}s]}\left(  \alpha\right)  $ where $I_{[.]}\left(  x\right)  =1$ if
$x\in\lbrack.]$ and $0$ otherwise ( Heaviside function ) and $d_{PF}$ is the
PF diameter. The parameter $q_{0}$ appearing in the angular part of
$\varepsilon_{pol}$ is the natural lattice twist that leads to the proper
geometric rotation of a PF around the tube axis. This parameter is lattice
type dependent and takes discrete values $2\pi/q_{0}=+3.4\mu m,+25\mu m,-6\mu
m$ for $N=12$, $13$ and $14$ PF MTs respectively \cite{Wade,Chretien}%
\cite{Ray}\cite{ChretienFuller}. We can estimate the prestrain $\varepsilon
_{PF}$ from the experimental value of the single switched PF's curvature
$\kappa_{PF}\approx\left(  250nm\right)  ^{-1}$ -measured in
Ref.\cite{Multistable Tub EM} on single taxol-stabilized PFs - to be
$\varepsilon_{PF}=d_{PF}\kappa_{PF}/2\approx10^{-2}.$ Collecting all energy
contributions together the total elastic +\ polymorphic energy of the MT is
then given by%
\begin{equation}
E_{MT}=\int\nolimits_{0}^{L}\left(  e_{el}+e_{trans}+e_{inter}\right)  ds.
\label{E_MTTotal}%
\end{equation}
The ground state within this model is determined by the interplay of the first
two terms $e_{el}$ and $e_{trans}.$ The last term $e_{inter}$ rules over
cooperativity and is responsible for the suppression of defects in the ideal
polymorphic order (cf. Fig. 7). A large value of the cooperativity constant
with $J>>k_{b}TL/b$ would imply a defect free lattice. However the presence of
the latter defects (and their motion) are a necessary ingredient for the
overall rearrangement of the helix as discussed later on.

\begin{figure}[ptbh]
\begin{center}
\includegraphics[
height=1.4512in, width=3.0735in ]{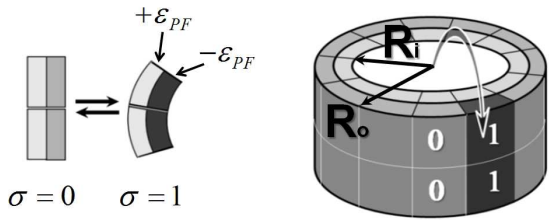}
\end{center}
\caption{Strains and deformations in the polymorphic tube model. Each tubulin
dimer can fluctuate between a straight state $\sigma=0$ and a curved state
$\sigma=1$ of intrinsic curvature $\kappa_{PF}$. The curved tubulin dimer
generates a positive prestrain $+\varepsilon_{PF}$ on its inner part and an
equal but negative prestrain $-\varepsilon_{PF}$ on its outer part with
strains related to observed dimer curvature via $\varepsilon_{PF}=(R_{o}%
-R_{i})\kappa_{PF}/2$.}%
\end{figure}

To understand the basic behavior of the ideal helical ground state without
defects - i.e. PFs are individually in a uniform state (either curved or
straight)- we initially restrict ourselves to the simplified energy density
$e=e_{el}+e_{trans}$. To investigate the MT geometry we first introduce two
reference frames (Fig. 2b). One is the material frame with base vectors
$(\vec{u}_{1},\vec{u}_{2},\vec{u}_{3})$ attached to the MT cross-section. The
other is an external fixed laboratory frame with base vectors $(\vec{u}%
_{x},\vec{u}_{y},\vec{u}_{z}).$ Putting the MT along the $\vec{u}_{z}$ axis
direction and considering small MT angular deflections we have $\vec{u}%
_{z}\approx\vec{u}_{3}$. In this case the two frames are simply related to
each other by a rotation transformation $R(s)$ given by the internal MT
lattice twist $q_{0}$, such that $(\vec{u}_{x},\vec{u}_{y})=R(s)(\vec{u}%
_{1},\vec{u}_{2})$ with
\begin{equation}
R(s)=\left(
\begin{array}
[c]{cc}%
\cos q_{0}s & -\sin q_{0}s\\
\sin q_{0}s & \cos q_{0}s
\end{array}
\right)  \label{R(s)}%
\end{equation}
To rewrite $e$ in a more illuminating fashion, we define two important order
parameters at each MT cross-section. The first one of them is the
\textit{vectorial polymorphic order parameter }
\[
\vec{P}\left(  s\right)  =\sum_{n=1}^{N}\left(  \vec{u}_{1}\cos\dfrac{2\pi
n}{N}+\vec{u}_{2}\sin\frac{2\pi n}{N}\right)  \sigma_{n}\left(  s\right)
\]
Physically, $\vec{P}$ - a 2D vector at each local material frame section
attached to the MT (cf Fig. 2b)- describes the asymmetry of distribution - a
kind of "polarization" - of the dimer states. It acquires a non zero value
only in the case when the curved and non-curved states are azimuthally
separated on opposite MT\ sides. For instance the "all-straight" or
"all-curved" PF state correspond both to the same value $\vec{P}=0$. Besides
the vector $\vec{P}$ we need to define a second (scalar) quantity
\[
M\left(  s\right)  =\sum_{n=1}^{N}\sigma_{n}\left(  s\right)
\]
which counts the total number of dimers in the curved state at cross-section
$s$ - or in the Ising model terminology : the "magnetization". After
integration of Eq. \ref{e_el} over the cross-section and some algebra the
energy density $e=e_{el}+e_{trans}$ can be written in a more appealing form:
\begin{equation}
e=\frac{B}{2}\left[  \left(  \vec{\kappa}-\vec{\kappa}_{pol}\right)
^{2}+\kappa_{1}^{2}\left(  \frac{\pi}{N}\gamma M-\sin^{2}(\pi/N)\vec{P}%
^{2}\right)  \right]  \label{EnergyPandM}%
\end{equation}
with the elastic bending modulus $B=\frac{Y\pi}{4}\left(  R_{o}^{4}-R_{i}%
^{4}\right)  $, with $\kappa_{1}=\frac{\left(  R_{0}-R_{1}\right)  ^{2}}%
{\pi\left(  R_{o}^{2}+R_{i}^{2}\right)  }\kappa_{FP}$ and a dimensionless
parameter
\begin{equation}
\gamma=\frac{\kappa_{PF}}{\kappa_{1}}-\frac{2N\Delta G}{bB\kappa_{1}^{2}}
\label{gamma}%
\end{equation}
For small deviations of the tube axis from the $\vec{u}_{z}$ direction, the
polymorphic curvature vector $\vec{\kappa}_{pol}$ in the external coordinate
frame $\left(  \vec{u}_{x},\vec{u}_{y}\right)  $ is\ related to $\vec{P}$ (in
the internal frame $\left(  \vec{u}_{1},\vec{u}_{2}\right)  $) via the
transformation in Eq. \ref{R(s)}:%

\begin{equation}
\vec{\kappa}_{pol}=cR(s)\vec{P} \label{Rot}%
\end{equation}
with $c=\sin\left(  \pi/N\right)  \kappa_{1}$ a geometric proportionality constant.

\subsection{Phase Diagram}

In the absence of defects the energy expression Eq. \ref{EnergyPandM} allows
us to determine the conformational ground state of a polymorphic MT. To this
end, we first resort to one further small simplification and make the "single
block ansatz", i.e. at each cross-section we assume only a single continuous
block of $p$ switched PFs. This ansatz was previously used by Calladine and
has been proven very useful in modelling bacterial flagellin polymorphic
states \cite{Asakura,Calladine}. In the ground-state configuration the
curvature is given by $\vec{\kappa}=\vec{\kappa}_{pol}(p)$ whose absolute
value obtained from Eq. \ref{Rot} is $\kappa_{pol}\left(  p\right)
=\kappa_{1}\sin\left(  \pi p/N\right)  $. The optimal switched block size
$p=p^{\ast}$ can be determined by minimizing the second term in Eq.
\ref{EnergyPandM} which within this ansatz becomes%

\begin{equation}
e=\frac{B\kappa_{1}^{2}}{2}\left(  \gamma\frac{\pi}{N}p-\sin^{2}\left(
\frac{\pi}{N}p\right)  \right)  \label{EnergyL}%
\end{equation}
This gives rise to an interesting MT phase behavior (cf. Fig. 6). The latter
only depends on the polymorphic-elastic competition parameter $\gamma$ from
Eq. \ref{gamma}\ -that measures the ratio between polymorphic energy of
tubulin switching and the purely elastic cost of this transition.

For $\gamma<-1$ the chemical switching potential $\Delta G$ strongly dominates
the elastic energy cost $bB\kappa_{1}^{2}$. Therefore switching is highly
favorable and all the PFs will be found in the state $\sigma=1$. This gives
rise to a straight but highly prestrained configuration.

Analogously for $\gamma>1$ the bending energy contribution is too costly for
PFs to switch at all. Therefore in this regime the PFs are all in the straight
state with $\sigma=0$ and the MT is consequently straight as well.

For $-1<\gamma<1$ the situation is more interesting. In this interval we have
a coexistence of two locally (meta) stable MT conformations : the\ straight
tube (prestressed or not - depending on the sign of $\gamma$)\ and a curved
lattice state with $p^{\ast}$ switched protofilaments. For $-\overline{\gamma
}<\gamma<\overline{\gamma}$ with $\overline{\gamma}\approx0.72$ the curved
lattice state is the absolute energy minimum and the straight state is only
metastable. Therefore in this regime and in the absence of twist $\left(
q_{0}=0\right)  $ the ground state configuration of the whole tube would be a
simple circular arc section (cf. Fig. 2b). On the other hand, the ground state
of a microtubule bearing natural lattice twist $q_{0}\neq0$ will be helical
(cf. Fig. 4).

\begin{figure}[ptbh]
\begin{center}
\includegraphics[
height=2.3981in, width=3.237in ]{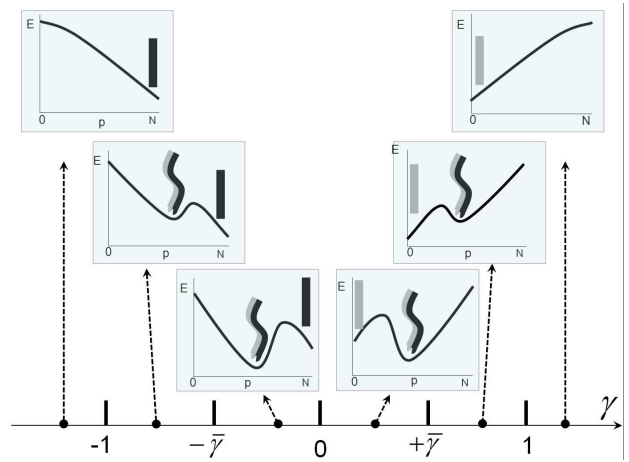}
\end{center}
\caption{The phase diagram of the polymorphic microtubule model as function of
the polymorphic-elastic interaction parameter $\gamma$ from Eq. \ref{gamma}.
Depending of the magnitude and sign of $\gamma$ the microtubule can be either
in an "all PF switched state" (black), "no PF switched state" (grey) or in a
mixed curved-helical state. Only the latter "mixed state" will display net
curvature and lead to an observable helical appearance.}%
\end{figure}

It is easy to see that a stable helical state as observed in \cite{Venier} is
only possible for a switching ratio $p^{\ast}/N\in\lbrack1/4,3/4]$. This
together with $\kappa_{PF}=1/250nm$ from \cite{Multistable Tub EM} and
$\kappa_{pol}\left(  p^{\ast}\right)  =\kappa_{1}\left\vert \sin\left(  \pi
p^{\ast}/N\right)  \right\vert $ provides us with a direct estimate of the
radius of curvature $\kappa_{pol}^{-1}\approx9-14\mu m$. Very strikingly the
latter range reproduces closely the observed MT helical curvatures as
estimated from Venier at al. work \cite{Venier} $\kappa_{measured}^{-1}%
\approx11\mu m$ - lending strong support to the model. Furthermore, taking the
helical state stability as an empirical fact (implying that $\left\vert
\gamma\right\vert <0.72$) and assuming a typical protein Young modulus
$Y\approx1-10GPa,$ allows us a simple estimate of the transition energy per
monomer $\Delta G\approx+1.1$ to $+11kT$ $-$ a reasonable range for a soft
biological object.

In general, the full energy expression -including the cooperativity energy
term- Eq. \ref{E_MTTotal}\ gives rise to a very complex behavior. Here we will
focus on some basic new phenomena. It turns out that the most remarkable
deviation from standard wormlike chain (WLC) behavior arises from the
fluctuation dynamics of the polymorphic order parameter's angular phase that
we consider in the following.

\subsection{Polymorphic phase fluctuations}

Here we introduce a sightly different phenomenological model that simplifies
the study of Eq. \ref{E_MTTotal} while still reflecting important aspects and
physical properties of it. In this section we assume as before that the
helical state is the ground state and consider now the effect of the
fluctuations around it. To this end we decompose the polymorphic order
parameter
\[
\overrightarrow{P}\left(  s\right)  =P\left(  s\right)  \left[  \cos\left(
\phi\left(  s\right)  \right)  \vec{u}_{1}+\sin\left(  \phi\left(  s\right)
\right)  \vec{u}_{2}\right]
\]
with $P\left(  s\right)  $ being the \textquotedblright polymorphic
amplitude\textquotedblright\ and $\phi\left(  s\right)  $ the
\textquotedblright polymorphic phase variable\textquotedblright. The latter
one determines the orientation of the switched block tubulin dimers at each
cross section with respect to the MT material frame. From Eq. \ref{Rot}, the
centerline curvature with respect to a fixed external $\left(  \vec{u}%
_{x},\vec{u}_{y}\right)  $ frame is then:%
\begin{equation}
\overrightarrow{\kappa}_{pol}(s)=\kappa_{0}\left[  \cos\left(  q_{0}%
s+\phi\left(  s\right)  \right)  \vec{u}_{x}+\sin\left(  q_{0}s+\phi\left(
s\right)  \right)  \vec{u}_{y}\right]  \label{kappaPol}%
\end{equation}
with $\kappa_{0}=cP(s)$.

In general both the polymorphic phase $\phi$ and amplitude $P$ can fluctuate
along the MT contour and give contributions to the polymorphic energy. The
phase fluctuations are induced by creation and motion of polymorphic "double
defects", cf. Fig. 7. The double defect - a kind of "polymorphic dislocation"
- that can be either left or right handed - maintains the number of switched
protofilaments constant while reorienting the direction of curvature. At zero
temperature the lattice would be defect-free, $\phi$ would be constant and the
polymorphic order parameter $\vec{P}$ would strictly follow the lattice twist.
The phase change $\phi^{\prime}\equiv d\phi/ds$ will deviate from zero if on
relevant length scales there are enough polymorphic double defects to allow
for a reorientation of the polymorphic order parameter away from optimum. The
double defects carry only a limited local energy cost $\Delta E=2J$ per defect
and can be easily thermally excited if $J\lesssim k_{B}T$. In the
approximation of a large number of PFs, assuming that $\phi$ can change
continuously we can write the phase contribution to the energy as
\begin{equation}
E_{pol}(\phi)=\frac{C_{\phi}}{2}\int_{0}^{L}ds\phi^{\prime2} \label{Ephi}%
\end{equation}
with the polymorphic phase stiffness $C_{\phi}\approx k_{B}T\frac{N^{2}b}%
{8\pi^{2}}\left(  2+e^{2J/k_{B}T}\right)  $ which can be related to the
density of double defects with energy $2J\;($cf. Fig. 7 and Appendix A),
giving rise to a new length scale - the \textit{polymorphic phase coherence
length} $l_{\phi}=C_{\phi}/k_{B}T.$ For MTs shorter than $l_{\phi}$ we will
observe coherent helices while on longer length scales the helix softens
significantly and looses eventually its helical appearance.

In contrast to the just discussed polymorphic dislocations which can be easily
thermally excited, the variation of the polymorphic amplitude $P$, i.e, change
of the number of switched PFs is more energetically costly. Any deviation of
$P$ from its optimum state $P^{\ast}$ (given by the phase diagram) is
associated with an energy cost $E\propto(\left\vert P\right\vert -\left\vert
P^{\ast}\right\vert ))^{2}\cdot l$ proportional to the length $l$ of the
region in the unfavorable state, cf. Fig. 7 (see Appendix B). Therefore we
conclude that on large enough scales the polymorphic phase fluctuations will
be the dominant effect. Based on this and on the observation of stable helical
states \cite{Venier} we will in the following assume $P=const.$

\begin{figure}[ptbh]
\begin{center}
\includegraphics[
height=1.8204in, width=2.9525in ]{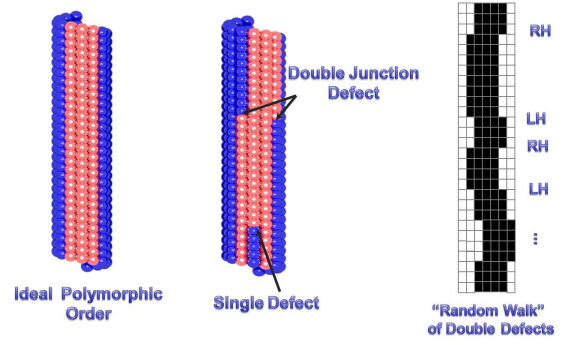}
\end{center}
\caption{Defects in ideal polymorphic order soften the helical states and give
rise to "polymorphic dynamics". Single defects carry a cost proportional to
their length, double defects only a local energy contribution. The coexistence
of left- and right handed defects (LH and RH) along the length leads to a
"random walk" of the polymorphic curvature direction and in turn to an
effective torsional deformation.}%
\end{figure}

For small deflections around the $z$-axis, the unit vector tangent to the MT's
centerline is approximately given by $\overrightarrow{t}\approx(\theta
_{x},\theta_{y},1)$\ in the laboratory frame $\left(  \vec{u}_{x},\vec{u}%
_{y},\vec{u}_{z}\right)  $\ where $\overrightarrow{\theta}=\left(  \theta
_{x},\theta_{y}\right)  $\ are the centerline deflection angles in x/y
direction. The global centerline curvature $\overrightarrow{\kappa
}=d\overrightarrow{t}/ds$\ can then be approximated as $\overrightarrow
{\kappa}\approx\left(  \theta_{x}^{\prime},\theta_{y}^{\prime},0\right)  $ and
the total MT energy can be written as follows:\
\begin{equation}
E_{tot}=E_{pol}\left(  \phi\right)  +E_{el}\left(  \theta,\phi\right)  .
\label{ET}%
\end{equation}
The first energy term is the polymorphic phase contribution Eq. \ref{Ephi}.
The second term is the elastic bending energy%
\begin{equation}
E_{el}\left(  \theta,\phi\right)  =\frac{B}{2}\int_{0}^{L}(\overrightarrow
{\theta^{\prime}}-\overrightarrow{\kappa}_{pol})^{2}ds. \label{Eel}%
\end{equation}
From Eqs. \ref{Ephi}-\ref{Eel}, we see that the zero-temperature ground state
corresponds to $\phi=const$ and $\overrightarrow{\theta^{\prime}%
}=\overrightarrow{\kappa}_{pol}$ - that is to a defect-free helix with a pitch
given by the natural lattice twist $q_{0}.$ At finite temperature, both
elastic and polymorphic fluctuations will be excited so that the curvature can
be decomposed as $\overrightarrow{\theta^{\prime}}=\overrightarrow{\kappa
}_{pol}+\overrightarrow{\theta^{\prime}}_{el}$ with $\overrightarrow
{\theta^{\prime}}_{el}$ the purely elastic contribution. This gives rise to a
helical MT shape described by the curvature $\overrightarrow{\kappa}%
_{pol}+\overrightarrow{\theta^{\prime}}_{el}$ and torsion $\tau\sim
\phi^{\prime}+q_{0}.$ The MT lateral displacements away from\ the $z$ axis can
be written as $\overrightarrow{\rho}(s)=(x(s),y(s))=\int_{0}^{s}\left(
\theta_{x}(s^{\prime})),\theta_{y}(s^{\prime}))\right)  ds^{\prime}$ that for
small deflections decouples into elastic and polymorphic displacements
$\overrightarrow{\rho}(s)=\overrightarrow{\rho}_{pol}+\overrightarrow{\rho
}_{el}$, where $\overrightarrow{\rho}_{el}\approx\int_{0}^{s}\overrightarrow
{\theta}_{el}(s^{\prime})ds^{\prime}$ and $\overrightarrow{\rho}_{pol}%
\approx\int_{0}^{s}\overrightarrow{\theta}_{pol}(s^{\prime})ds^{\prime}.$ The
latter can also be written from Eq. \ref{kappaPol} as
\begin{align}
\overrightarrow{\rho}_{pol}(s)  &  =\kappa_{0}\int_{0}^{s}ds^{\prime}\int
_{0}^{s^{\prime}}d\widetilde{s}\left(  \cos\left(  q_{0}\widetilde{s}%
+\phi\left(  \widetilde{s}\right)  \right)  \overrightarrow{e}_{x}\right.
\nonumber\\
&  \left.  +\sin\left(  q_{0}\widetilde{s}+\phi\left(  \widetilde{s}\right)
\right)  \overrightarrow{e}_{y}\right)  \label{Polydeplacement}%
\end{align}
In Fig. 8, snapshots of configurations of clamped MT obtained from Monte Carlo
simulations are plotted for different concentrations of double defects (i.e.
different values of $l_{\phi}$) with twist and no twist. It is interesting to
remark at this point that based on the symmetry in the problem any MT
configuration can be rotated around the $z-$axis with no energy cost. This
seemingly trivial feature - the energetic degeneracy- is in fact the most
distinctive and unusual property of a polymorphic chain. We consider the
consequences in the following.

\begin{figure}[ptbh]
\begin{center}
\includegraphics[
height=1.8213in, width=3.2387in ]{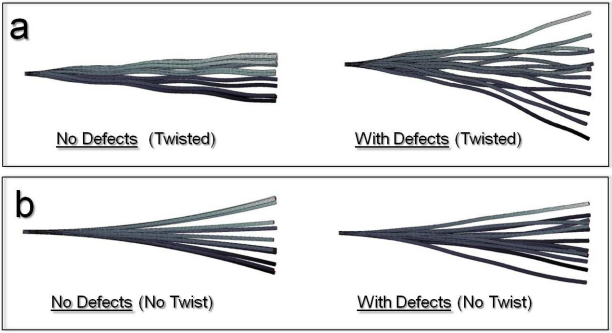}
\end{center}
\caption{The conical "wobbling" motion due to the rotational energy degeneracy
of the polymorphic MT model : Snapshots of Monte-Carlo simulated lattice
states. (a) Twisted MTs free of defects $L/l_{\phi}\ll1$ and with numerous
defects $L/l_{\phi}>1$. For larger number of defects, the helix looses its
coherent look. (b) For non-twisted MTs the movement has a typical parabolic
"trumpet-like" shape. }%
\end{figure}

\subsection{The Wobbling Mode}

By construction a polymorphic chain as we describe it here has a $N$ fold
symmetry. Therefore, there are $N$ different helical states of different
orientations with the same energy i.e. $N$ grounds states. This energy
degeneracy is also reflected in the continuum model (where the $N$ fold
symmetry is approximated as continuous) by the rotational invariance of
$E_{pol}\left(  \phi\right)  $ (which depends only on $\phi^{\prime}$). The
broken cylindrical to helical symmetry of the straight state is then restored
by the presence of a \textquotedblright Goldstone mode\textquotedblright%
\ $\phi\rightarrow\phi+\phi_{0}$ consisting of a rotation of $P$ by an
arbitrary angle $\phi_{0}$ in the material frame (cf. Figs. 2 and 3). This
mode comes energetically for free and leads to dramatic effects on chain's
fluctuations. For instance for a MT clamped at one end, this symmetry implies
that the MT will randomly rotate much like a rigid rotor as shown in Fig. 4.
Note however that this rotation will still be associated with a certain
dissipation as the system has to go over energy barriers between two helical
states. This barrier due to the flipping of lattice states can be overcome at
nonzero temperatures by the creation of double defects which diffusively
propagate along the MT and eventually angularly reorient the polymorphic order
parameter $\vec{P}$. These dynamic phenomena and the dissipation mechanisms
will be discussed in a later section.

In summary, the zero-energy mode that we will also call the \textquotedblright
wobbling mode\textquotedblright\ is an inherent feature of a helically
polymorphic filament and as we will see now could be the culprit causing
anomalous fluctuations of clamped MTs.

\subsection{Persistence Length Anomalies}

Among several definitions of the persistence length, we consider here - for
direct comparison with clamped MTs experiments \cite{Pampaloni,Taute} - the
"lateral fluctuation persistence length" defined as
\begin{equation}
l_{p}^{\ast}(s)=\left(  2/3\right)  s^{3}/V(s) \label{LpStar}%
\end{equation}
with $V(s)=\left\langle \rho\left(  s\right)  ^{2}\right\rangle -\left\langle
\rho\left(  s\right)  \right\rangle ^{2}$ the variance of $\rho^{2}%
=x^{2}+y^{2}$ - the transverse displacement at position $s$ and $\left\langle
..\right\rangle $ the ensemble average. We assume as in experiments
\cite{Pampaloni,Taute} a rigid clamping point at the position $s=0$ preventing
the microtubule from translating and rotating at that point.

For an ideal semiflexible wormlike chain we expect that the persistence length
$l_{p}^{\ast}=l_{B}$ is a position independent and definition invariant
quantity equal to the bending persistence length $l_{B}=B/k_{B}T$. (For
another more classical definition of the persistence length - coming from the
tangent tangent correlation function, see also Appendix C). However for a
polymorphic chain the strict equivalence of $l_{p}^{\ast}$ and $l_{B}$ is not
correct. To see that, we can decompose the polymorphic and elastic
fluctuations $\overrightarrow{\rho}(s)=\overrightarrow{\rho}_{pol}%
+\overrightarrow{\rho}_{el}$. Inserting this in Eq. \ref{LpStar} and taking
into account that for small deflections the two components decouple
$\left\langle \overrightarrow{\rho}_{pol}\overrightarrow{\rho}_{el}%
\right\rangle =0$ leads us to the following relation for the persistence
length
\begin{equation}
l_{p}^{\ast}=\left(  l_{pol}^{-1}+l_{B}^{-1}\right)  ^{-1} \label{lptotal}%
\end{equation}
with the polymorphic persistence length given by $l_{pol}=\left(  2/3\right)
s^{3}/V_{pol}$ and $V_{pol}=\left\langle \rho_{pol}{}^{2}\right\rangle
-\left\langle \rho_{pol}\right\rangle ^{2}$. The average $\left\langle
..\right\rangle $ is now performed over the phase $\phi$ governed by the
energy Eq. $\ref{Ephi}$. More precisely, the average of any arbitrary
functional $A\left[  \phi\right]  $ of the polymorphic phase can be performed
by first selecting one of the equivalent ground states denoted $\phi_{0}$
characterized by $\phi(0)=\phi_{0}$ and performing the average over the
polymorphic angle distribution Eq. $\ref{Ephi}$ around the chosen ground
state, i.e.,
\begin{equation}
\left\langle A\left[  \phi\right]  \right\rangle |_{\phi_{0}}=\frac{1}{Z}%
\int\emph{D}\tilde{\phi}A\left[  \tilde{\phi}+\phi_{0}\right]  \exp
(-\frac{l_{\phi}}{2}\int_{0}^{L}ds\tilde{\phi}^{\prime2}) \label{Aphi}%
\end{equation}
where $\tilde{\phi}=\phi-\phi_{0}$ (and thus $\tilde{\phi}(0)=0$) and
$Z=\int\emph{D}\tilde{\phi}\exp(-\frac{l_{\phi}}{2}\int_{0}^{L}ds\tilde{\phi
}^{\prime2})$ is the partition function. In a second step - for a freely
rotating polymorphic phase- we integrate over the rotational zero mode
$\phi_{0}$ : $\left\langle A\left[  \phi\right]  \right\rangle =\frac{1}{2\pi
}\int\left\langle A\left[  \phi\right]  \right\rangle |_{\phi_{0}}d\phi_{0}$.
This operation correctly takes into account the phase fluctuations over (and
around) all equivalent ground states related by the transform $\phi
\rightarrow\tilde{\phi}+\phi_{0}$. The rotational symmetry around the z-axis
(integration on $\phi_{0}$) readily implies $\left\langle \rho_{pol}\left(
s\right)  \right\rangle =0$ and $\left\langle x_{pol}^{2}\right\rangle
=\left\langle y_{pol}^{2}\right\rangle .$ Therefore the polymorphic
persistence length can be written as
\begin{equation}
l_{pol}(s)=\left(  1/3\right)  s^{3}/\left\langle y_{pol}^{2}(s)\right\rangle
. \label{lpol}%
\end{equation}
with
\begin{equation}
\left\langle y_{pol}^{2}(s)\right\rangle =\int_{0}^{2\pi}\frac{d\phi_{0}}%
{2\pi}\int_{0}^{s}\int_{0}^{s}\left\langle \theta_{y,pol}(s_{1})\theta
_{y,pol}(s_{2})\right\rangle |_{\phi_{0}}ds_{1}ds_{2} \label{ypolsquare}%
\end{equation}
whose computation (for details cf. Appendix C) leads to the following mean
square displacement
\begin{align}
\left\langle y_{pol}(s)^{2}\right\rangle  &  =\frac{2\kappa_{0}^{2}l_{\phi}%
}{3(1+4l_{\phi}^{2}q_{0}^{2})^{4}}\left\{  P_{1}\left(  s\right)
-e^{-\frac{s}{2l_{\phi}}}P_{2}\left(  s\right)  \cos\left(  q_{0}s\right)
\right. \nonumber\\
&  \left.  -e^{-\frac{s}{2l_{\phi}}}P_{3}\left(  s\right)  \sin(q_{0}%
s)\right\}  \label{y2pol}%
\end{align}
where $P_{i}(s)$ are polynomial functions given in Appendix C.\

A typical curve of $l_{p}^{\ast}$ vs $s$ is provided in Fig. 9. In general it
shows three different characteristic regimes denoted I, II and III in the
figure :

I.\ At short distances to the attachment point $s<s_{\min}\approx\pi/q_{0}$
(half the polymorphic wavelength) the total persistence length can be
approximately given by%

\begin{equation}
l_{p}^{\ast}\approx l_{B}-\frac{3\kappa_{0}^{2}l_{B}^{2}s}{8}
\label{lp short L}%
\end{equation}
In the limit of very short distances $s<<l_{B}^{-1}\kappa_{0}^{-2},$ the
polymorphic fluctuations become negligible and are completely dominated by
purely "classical" semiflexible chain fluctuations. Not surprisingly the
persistence length coincides then with the classical bending persistence
length $l_{p}^{\ast}\left(  0\right)  =l_{B}$. Starting from $l_{B}$,
polymorphic fluctuations begin to contribute reducing $l_{p}^{\ast}$ that
attains a global minimum at $s_{\min}\approx\pi/q_{0}$.

II. For intermediate length values $s_{\min}<s<$ $l_{\phi}$, the total
persistence length displays a non-monotonic oscillatory behavior around a
nearly linearly growing average%
\begin{equation}
l_{p}^{\ast}\left(  s\right)  \approx\frac{2}{3}\frac{q_{0}^{2}}{\kappa
_{0}^{2}}s+\frac{4}{3}\frac{q_{0}}{\kappa_{0}^{2}}\sin\left(  q_{0}s\right)
\label{lp longue L}%
\end{equation}
This result is worth deeper understanding. A moment of thinking reveals that
the oscillatory part with wavelength $2\pi/q_{0}$ is related to the helicity
of the ground state. At the same time the linear growth $l_{p}^{\ast}\left(
s\right)  \propto\alpha^{2}s$ can be associated with the roughly conical
rotation of the clamped chain (wobbling mode) which acts as an effective
\textquotedblright rotational hinge\textquotedblright\ at the attachment
point, cf. Fig. 4. The sinusoidally modulated rotation cone which builds an
approximate envelope for the chains' motion has an opening angle $\alpha$
which is related to the geometric features of the helix $\alpha=2\kappa
_{0}q_{0}^{-1}$.

III. Finally for very large distances from the attachment point $s\gg l_{\phi
}$ we expect to recover classical results of a semiflexible chain again.
Indeed in this asymptotic regime the effective persistence length reaches
saturation with a renormalized constant value $l_{p}^{\ast}\left(
\infty\right)  =1/\left(  l_{pol}^{-1}+l_{B}^{-1}\right)  $\ where
\begin{equation}
l_{pol}=2l_{\phi}q_{0}^{2}\kappa_{0}^{-2}+\frac{1}{2}\kappa_{0}^{-2}l_{\phi
}^{-1} \label{lpinfinity}%
\end{equation}
Intuitively the helix looses then its "coherent nature" - due to strong
variations of $\phi^{\prime}$ and elastic fluctuations $\theta_{el}$ - and the
collective rigid rotational (\textquotedblright conical\textquotedblright)
motion is finally replaced by an uncorrelated segment movement. Not
surprisingly the persistence length becomes then length independent again.
Curves for different values of $l_{\phi}$ are provided in Fig. 10.

\subsubsection{Untwisted MTs}

While there are no completely twist-free MTs and every lattice will have
generically a small twist, one can still formally study the interesting
limiting case $q_{0}=0$. Note that the large estimated pitch of $13$PF MTs is
finite and in the range of $\gtrsim25\mu m$
\cite{Wade,Chretien,Ray,ChretienFuller} while often assumed to be
approximately infinite. In such an ideal case the theory still applies,
however the overall behavior of $l_{p}^{\ast}\left(  L\right)  $ will
substantially change and become much less consistent with the linear scaling
found in experiments. While for $q_{0}L>1$ the chain in leading order moves on
a linear cone (with fixed opening angle $\alpha$) for $q_{0}L<<1$ while still
$L<<l_{\phi}$ the \textquotedblleft wobbling\textquotedblright\ motion is
happening on a quadratic cone (a \textquotedblleft trumpet
shaped\textquotedblright\ cone, cf. Fig. 8b). More precisely for the
exceptional case of untwisted MTs the polymorphic part of the lateral
fluctuations Eq. \ref{y2pol} behaves as $\left\langle y_{pol}(s)^{2}%
\right\rangle =2/3\kappa_{0}^{2}l_{\phi}\left(  P_{1}-e^{-\frac{s}{2l_{\phi}}%
}P_{2}\right)  $ with $P_{1}(s)=24l_{\phi}^{3}-3l_{\phi}s^{2}+s^{3}$,
$P_{2}(s)=24l_{\phi}^{3}+12l_{\phi}^{2}s{\small .}$ Therefore for short MTs
$L<<l_{\phi},$ the lateral fluctuations grow with the length in fourth power
$\left\langle y_{pol}(L)^{2}\right\rangle \approx\kappa_{0}^{2}{\small L}%
^{4}/8$, whereas for long ones $L>>l_{\phi},$ the deviation grows cubically,
$\left\langle y_{pol}(L)^{2}\right\rangle ={\small 2/3}\kappa_{0}^{2}l_{\phi
}{\small L}^{3}.$ From this, the persistence length has consequently two
typical regimes. For $L<<l_{\phi}$ we deduce from Eqs. \ref{lptotal} and
\ref{lpol} that $l_{p}^{\ast}\left(  L\right)  \approx\left(  l_{B}%
^{-1}+3/8\kappa_{0}^{2}L\right)  ^{-1}$. The latter expression implies that
long, untwisted MTs appear increasingly softer with growing length, and the
effective persistence length decays inversely $l_{p}^{\ast}\propto1/L$ for
$L>>1/(3/8\kappa_{0}^{2}l_{B})$ and reaches a limiting value $l_{p}^{\ast
}=l_{B}/\left(  1+2l_{B}l_{\phi}\kappa_{0}^{2}\right)  $ for $L>>l_{\phi}.$
For a MT of $L\sim10-20
\mu m$ we would expect $l_{p}^{\ast}\sim10-20\mu m$ - a value $2$ orders of
magnitude smaller than observed in \cite{Pampaloni,Taute}. This decreasing
behavior is in contrast with observations $l_{p}^{\ast}\propto L$ - leading us
to the conclusion that the $0$-twist MTs does not constitute a significant
portion of the experimental data \cite{Pampaloni,Taute} and twist is
necessarily required for the growth of $l_{p}^{\ast}$ with $L.$

Having developed some static consequences of the polymorphic MT we now turn to
its dynamical aspects.

\begin{figure}[ptb]
\begin{center}
\includegraphics[
height=2.2061in ]{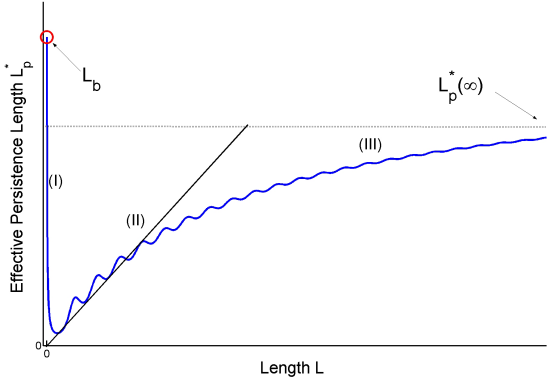}
\end{center}
\caption{A typical shape of the effective persistence length $l_{p}^{\ast
}\left(  L\right)  $ for a clamped microtubule as obtained from Eqs.
\ref{lptotal}, \ref{lpol} and \ref{y2pol}. Most generically the curve displays
three different regimes: (I) An initial rapidly decreasing regime where
polymorphic effects become more effective with growing $L$ (softening the
chain), (II) a linearly-growing oscillatory regime - corresponding to the
coherent wobbling movement of the clamped microtubule, (III) an asymptotic
plateau regime, where the helix progressively loses its coherence with growing
$L.$ In this regime, the behavior tends to that of a classical semiflexible
chain yet with a renormalized effective persistence length given by cf. Eq.
\ref{lpinfinity}.}%
\end{figure}

\begin{figure}[ptb]
\begin{center}
\includegraphics[
height=2.0773in, width=3.3157in ]{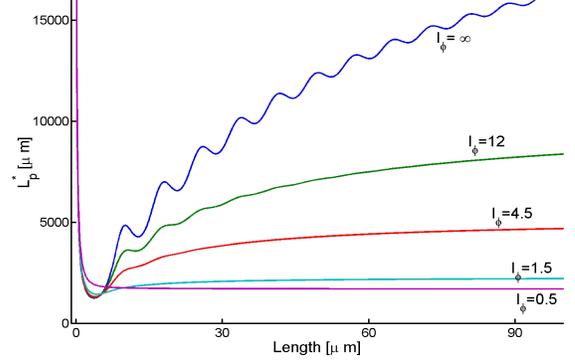}
\end{center}
\caption{Comparison of theoretical persistence lengths for different values of
the polymorphic phase coherence length $l_{\phi},$ ($l_{B}=25mm,$ $\kappa
_{0}=0.03\mu m^{-1}$ and $q_{0}=0.8\mu m^{-1}$). For $l_{\phi}=\infty,$ the MT
is a (defect free) coherent helix performing the "wobbling motion"(as in Figs.
4 and 8a left panel) . The plateau regime - where elastic fluctuations become
dominant over polymorphic- is reached for very long MT only (not seen in the
Fig.). Finite $l_{\phi}$ reduces the coherent wobbling motion - shortens
region (II) of Fig. 9 - and the plateau regime is reached earlier with
decreasing $l_{\phi}$.}%
\end{figure}

\subsection{Polymorphic Phase Dynamics}

To describe the MT fluctuation dynamics we consider the total dissipation
functional $P_{diss}=P_{ext}+P_{int}$ which is composed of an internal
dissipation $P_{int}=\frac{1}{2}\xi_{int}\int\dot{\phi}^{2}ds$ (with
$\dot{\phi}\equiv d\phi/dt$) coming from the flipping of lattice states and an
external hydrodynamic\ dissipation $P_{ext}=\frac{1}{2}\xi_{\perp}\int
\overset{\cdot}{\rho}^{2}ds$ associated with the time variation of the MT
deflection $\overrightarrow{\rho}\left(  s,t\right)  =(x(s,t),y(s,t))$. We
assume that the friction constant (per unit length) $\xi_{\perp}$ of the
helical MT is approximately the friction constant $\xi_{\perp}=4\pi
\eta/\left(  \ln\left(  2L/R\right)  -1/2\right)  $ of a long slender body of
length $L,$ (small)\ radius $R<<L$ moving in a fluid with viscosity $\eta$ at
low Reynolds numbers. The time evolution equation of the phase variable
$\phi\left(  s,t\right)  $ and the lateral displacement $y\left(  s,t\right)
$ (and $x\left(  s,t\right)  $) are given by the coupled Langevin equations
\begin{equation}
\frac{\delta E_{tot}}{\delta\phi}=-\frac{\delta P_{diss}}{\delta\dot{\phi}%
}+\Gamma_{\phi} \label{LangevinPhi}%
\end{equation}
and
\begin{equation}
\frac{\delta E_{tot}}{\delta y}=-\frac{\delta P_{diss}}{\delta\dot{y}}%
+\Gamma_{\rho}%
\end{equation}
with $\Gamma_{\phi/\rho}$ the thermal noise terms. In general the lateral
displacement $y\left(  s,t\right)  $ has contributions from both polymorphic
$y_{pol}(s,t)\approx\kappa_{0}\int_{0}^{s}ds^{\prime}\int_{0}^{s^{\prime}}%
\sin\left(  \phi(\tilde{s},t)+q_{0}s\right)  d\widetilde{s}$ and elastic
fluctuations $y_{el}(s,t)\approx\int_{0}^{s}\theta_{el}(s^{\prime
},t)ds^{\prime}$ and the dynamics is highly non-linear. In the regime
$L>>l_{\phi}$ where the helix looses its coherence one expects to retrieve the
dynamics of the usual semiflexible filament with $\tau\sim L^{4}$. However in
the opposite and physically more interesting regime $L<<l_{\phi}$ a new and
different dynamic behavior can be expected. As we learnt from the study of the
static case the effects of polymorphism become more pronounced at shorter
lengths. As we have seen in this regime, the dominant motion is the wobbling
rotation of a coherent helix on a cone where elastic fluctuations become
negligible compared to polymorphic ones i.e. $y(s,t)\approx y_{pol}(s,t)$. In
this regime few polymorphic defects $L<<l_{\phi}$ are present and the phase
can be approximated as $\phi\left(  s,t\right)  \approx\phi_{0}(t)+\delta
\phi\left(  s,t\right)  .$Using this decomposition with $\delta\phi\left(
s,t\right)  <<1$ we can expand $P_{ext}$ to leading order
\begin{equation}
P_{diss}\approx\frac{1}{2}L\left(  \xi_{int}+\xi_{ext}\right)  \dot{\phi}%
_{0}^{2}+O\left(  \delta\dot{\phi}^{2}\right)
\end{equation}
with a external friction constant $\xi_{ext}$ given by
\begin{equation}
\xi_{ext}=\frac{\xi_{\perp}\kappa_{0}^{2}}{q_{0}^{4}}(2\left(  1+\cos
Lq_{0}\right)  -4\frac{\sin Lq_{0}}{Lq_{0}}+\frac{q_{0}^{2}L^{2}}{3})
\label{Xiexternal}%
\end{equation}
The evolution of the zero mode $\phi_{0}\left(  t\right)  $ reduces from the
Langevin equation, Eq. \ref{LangevinPhi}, to $0=-\frac{\delta P_{diss}}%
{\delta\dot{\phi}_{0}}+\Gamma_{\phi}$ which leads to the equation of motion%

\begin{equation}
\frac{d}{dt}\phi_{0}\left(  t\right)  =\frac{1}{\xi_{tot}}L^{-1}\int_{0}%
^{L}\Gamma_{\phi}\left(  s,t\right)  ds \label{phiotime}%
\end{equation}
with a friction constant $\xi_{tot}=\xi_{int}+\xi_{ext}.$ Therefore $\phi
_{0}\left(  t\right)  $ satisfies the simple Langevin equation Eq.
\ref{phiotime} corresponding to a simple potential free Brownian motion with
its mean square displacement given by (see Appendix D for a more detailed
explanation)
\begin{equation}
\left\langle \left(  \phi_{0}\left(  t\right)  -\phi_{0}\left(  0\right)
\right)  ^{2}\right\rangle =\frac{2k_{B}T}{L\xi_{tot}}t. \label{MSDphio}%
\end{equation}
In this limit (wobbling mode dominant) we have a roughly rigid helix moving
randomly along a cone and all the physics is contained in the effective
friction coefficient $\xi_{tot}\left(  \xi_{int},\kappa_{0},q_{0},L\right)  $
and its dependence on the internal dissipation $\xi_{int},$ the helix
parameters $\kappa_{0},$ $q_{0}$ and the length $L$.

For later comparison with experiments we compute the longest relaxation time
$\tau$\ given by the auto-correlation function $\left\langle y_{pol}%
(s,0)y_{pol}(s,t)\right\rangle \propto e^{-t/\tau}$. Using Eq. \ref{MSDphio},
a short computation (Appendix D) leads to%

\begin{equation}
\left\langle y_{pol}(L,0)y_{pol}(L,t)\right\rangle =\left\langle y_{pol}%
^{2}(L)\right\rangle e^{-t/\tau(L)}%
\end{equation}
with $\left\langle y_{pol}^{2}(L)\right\rangle =\frac{\kappa_{0}^{2}}%
{q_{0}^{2}}\left(  \frac{L^{2}}{2}+\frac{1-\cos(q_{0}L)}{q_{0}^{2}}%
-\frac{L\sin(q_{0}L)}{q_{0}}\right)  $\ the static mean square displacement
Eq. \ref{y2pol} in the limit $L<<l_{\phi}$\ and with $\tau\left(  L\right)
=L\xi_{tot}/k_{B}T$ the longest relaxation time - proportional to the total
friction constant $\xi_{tot}$ - inheriting its length dependence in two
different length regimes. For very short lengths $L\ll l_{c}=\left(
3\xi_{int}\xi_{\perp}^{-1}\right)  ^{1/2}q_{0}\kappa_{0}^{-1}$ when the
hydrodynamic dissipation is entirely dominated by internal dissipation we have
a linear scaling
\begin{equation}
\tau\left(  L\right)  \approx L\xi_{int}/k_{B}T. \label{tau1}%
\end{equation}
For larger $L>l_{c}$ the wobbling movement through the fluid is the dominant
source of dissipation and%

\begin{equation}
\tau\left(  L\right)  \approx L\xi_{ext}(L)/k_{B}T. \label{Tau2}%
\end{equation}
with $\xi_{ext}(L)$ given by Eq. \ref{Xiexternal}. Note that the $L$
dependence of $\xi_{ext}$ relies also on the $L$ dependence of $\xi_{\perp
}(L)$ which for simplicity has been modelled as the friction of an ideal
slender tube moving in a liquid. A more precise (but difficult) determination
of $\xi_{\perp}$ could slightly change its variation with $L,$ although not
the general trends of $\xi_{ext}(L).$ In particular if we assume that
$\xi_{\perp}$ is $L$ independent and that $Lq_{0}\gg1$ we have in this regime
the scaling
\begin{equation}
\tau\left(  L\right)  \approx\frac{\xi_{\perp}}{3k_{B}T}\left(  \kappa
_{0}/q_{0}\right)  ^{2}L^{3}. \label{Tau3}%
\end{equation}
So in summary for $\tau$ we expect a cross-over from a linear, internal
dissipation dominated $L$ dependence at short lengths to a cubic length
dependence given by hydrodynamic friction alone.

\section{Comparison with Clamped MT Experiments}

The comparison with experiments \cite{Pampaloni,Taute} which measure lateral
fluctuations of clamped MTs reveals several interesting characteristics that
are in agreement with predictions, cf. Fig. 11. First, the predicted mean
linear growth of $l_{p}^{\ast}\left(  L\right)  $ agrees with experiments as a
single exponent fit $l_{p}^{\ast}\sim L^{\delta}$ of the data provides
$\delta=1.05$. Besides the linear growth the experimental data reveal a large
spread of $l_{p}^{\ast}$ data points which seems to grow approximately in
proportion to the length. This linearly growing experimental spread is likely
linked to the intrinsic spread of $q_{0}$ values of different MT lattice
populations \cite{Wade,Chretien,Ray,ChretienFuller}. MTs with different number
of protofilaments will display different lattice twists ranging from
$q_{0}\approx2\pi/3\mu m$ \ (for 12 PF\ MTs) to $q_{0}\lesssim2\pi/25\mu m$
\ (for 13 PF\ MTs). Keeping in mind the scaling $l_{p}^{\ast}\left(  s\right)
\propto q_{0}^{2}s$ (cf. Eq. \ref{lp longue L}) we would expect more than an
order of magnitude variation of measured $l_{p}^{\ast}$ while the slope
$l_{p}^{\ast}\left(  s\right)  /s$ should display a constant spread.

Second, the data of Taute et al. \cite{Taute} (Fig. 11, circle) indicate a
non-monotonic dependence with systematic trends over several consecutive data
points. This seems phenomenologically well captured by the oscillatory
behavior $l_{p}^{\ast}\left(  L\right)  $ from Eq. \ref{lp longue L}. On the
other hand, the data of Pampaloni et al. \cite{Pampaloni} are definitely much
more spread and don't allow such a clear conclusion. Therefore the
non-monotonicity of $l_{p}^{\ast}\left(  L\right)  $ is at present
experimentally difficult to infer from the two existing experimental data sets
taken together, however it is consistent with data within the error bars. As
mentioned the presence of different lattice populations (even within single
MTs) could give rise to a large spread of experimental data points and an
effective "washing out" of the non-monotonic behavior for different lattices
within the same statistics.

Remarkably the experimental data reveal that the large length plateau $s\gg
l_{\phi}$ where $l_{p}^{\ast}$ would become length independent is not reached
even for longest MTs ($\sim50\mu m$). This is in phenomenological agreement
with the theory - as based on the observed long coherent helices by Venier et
al \cite{Venier} it would imply a very long $l_{\phi}$. The absence of the
plateau in Pampaloni and Taute's data allows a lower estimate of the coherence
length: $l_{\phi}>55\mu m$ which in turn would imply a large coupling constant
$J>4k_{B}T$. The best comparison between theory and experiments (cf. Fig. 11)
gives $l_{B}=25mm$ corresponding to a rather high Young modulus $Y\approx9GPa$
higher than typically reported before
\cite{GITTES,Venier,Mickey,Felgner,Kukimoto,ActtiveMTBending,TAKASONEbuckling,Brangwyne,Janson}%
. However the present value is well within the range for proteins and protein
tubes with $Y\ $up to $19GPa$ are reported in literature \cite{15 GPa Tubes}.
The higher value of the bare Young modulus extracted form the present theory
should also not be a surprise as in previous models all MT conformational
fluctuations were interpreted as originating from bending deformations alone.
In our model - both elastic and polymorphic fluctuations contribute with the
latter being much softer and giving therefore dominant contribution.

The best fitting helix wave length $\lambda\approx7.5\mu m$ is close to the
expected $6\mu m$ corresponding to the twist
\cite{Wade,Chretien,Ray,ChretienFuller} of the $14$ PF MT population. This is
in agreement with the fact that, - in contrast to the in vivo situation- the
large pitch ($\lambda\approx25\mu m$) $13$PF MTs are likely underrepresented
in the data \cite{Pampaloni,Taute}. Indeed in vitro studies of
taxol-copolymerized MTs display a MT population consisting of a majority of
$14$ PFs ($61\%$) while $13$ PFs ($32\%$) are less represented
\cite{Wade,Chretien,Ray,ChretienFuller}. The in vitro conditions therefore
strongly shift the PF population away from the preferred low twist $13$ PF MT
towards the highly twisted $14$PF MT.

\begin{figure}[ptbh]
\begin{center}
\includegraphics[
height=2.8003in, ]{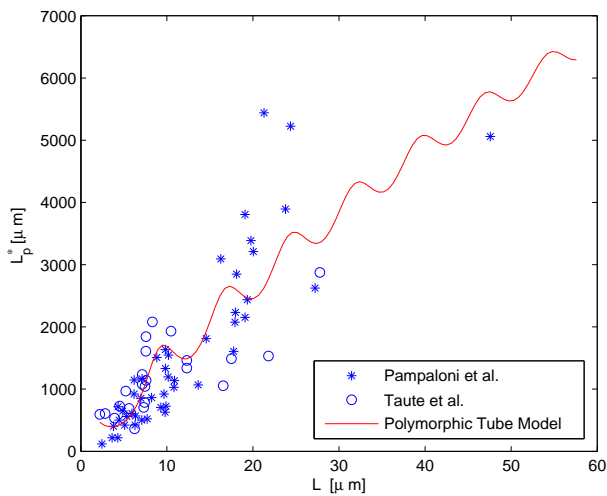}
\end{center}
\caption{Effective persistence length $l_{p}^{\ast}$ as a function of the
position from the attachment point along the clamped MT contour. The
experimental data (stars and circles) \cite{Pampaloni,Taute} and the
theoretical prediction with $l_{B}=25mm,$ $\lambda=7.5\mu m,$ $\kappa_{0}%
^{-1}=18\mu m,$ $q_{0}l_{\phi}\gg1$. }%
\label{persistence length}%
\end{figure}

The estimated $l_{B}$ is larger than in previous studies ($l_{B}\sim1-6mm)$
where however polymorphic fluctuations were neglected. This result leads us to
the conclusion that if polymorphism is partially suppressed one would measure
much larger effective $l_{p}$ - as in fact observed. For example in studies
where 2D slab geometry is used (MTs between 2 close glass slides) an effective
suppression of the 3 dimensional polymorphic helices or their reduced mobility
is expected. A typical observation in such cases is an extensive
\textquotedblleft intrinsic curvature\textquotedblright\ (of previously
unknown origin). Within our theory one could interpret this curvature as
pinned / quenched polymorphic helices prevented from free fluctuations by the
confinement. These effects could in general explain the dramatic variations of
measured $l_{p}$ values based on the presence/ absence of polymorphic
\textquotedblleft softening\textquotedblright\ in different experimental
setups and geometries.

Now, let us consider the clamped MT dynamics as investigated in \cite{Taute}.
A careful analysis of Taute et al's data reveals a peculiar scaling of the
longest relaxation time with the length. Indeed an independent single exponent
fit of the Taute et al data \cite{Taute} gives $\tau\propto L^{\alpha}$ with
$\alpha=2.9$ in the experimental range considered $2.2\mu m<L<28\mu m$. This
peculiar scaling can be understood as originating from hydrodynamic relaxation
of a "wobbling" polymorphic chain. Using only the previously best fitting
parameters of the static data (Fig. 11) $\left(  \kappa_{0}/q_{0}\right)
^{2}\approx4.\,8\times10^{-3}$ and $\eta=10^{-3}Pa\cdot s$ (water viscosity)
we find a remarkable correspondence between the theoretical prediction Eq.
\ref{Tau2}, i.e., $\tau_{th}\left(  L\right)  \approx L\xi_{ext}(L)/k_{B}T$
and data of \cite{Taute}, as shown in Fig.\ 12.\ This 0-parameter prediction
matches well the data for larger lengths. For scaling comparison, it is
interesting to compare the data with the approximate theoretical relaxation
time Eq. \ref{Tau3}, i.e., $\tau\left(  L\right)  \approx\frac{\xi_{\perp}%
}{3k_{B}T}\left(  \kappa_{0}/q_{0}\right)  ^{2}L^{3}$ (expression strictly
valid for $L\gg0.8$ $\mu m$) which has a scaling law in agreement with the
single exponent fit of the data. Considering that $\xi_{\perp}\approx
1.6\eta-2.3\eta$ is roughly length independent in the experimental $L$ range
one can assume $\xi_{\perp}\approx2\eta$ and obtains the prefactor $\tau
_{th}/L^{3}=7.9\times10^{14}s/m^{3}.$ Keeping in mind the simplicity of the
interpretation (and the lack of free parameters therein), this compares very
favorably with the best fit of experimental data slope $\tau_{fit}%
/L^{3}=6.\,25\times10^{14}s/m^{3}$ (cf. Fig. 12). Note that the approximate
value of $\tau_{th}$ seems to correspond a little better to the data but this
is likely of no physical significance at this level of approximation. Indeed
the neglected elastic modes other than the wobbling as well as the
approximation of the hydrodynamic friction should change the details of the
$L$ dependence of $\xi_{ext}$ and thus of $\tau_{th}$ (but not the general
trends). Without a more precise computation of the dynamics we can be
satisfied with the rather astonishing agreement between theory and experiment
for long MTs. This leads us again to the conclusion that in these experiments
long MTs behave as almost rigid helical polymorphic rotors whose motion is
dominated by the zero energy \textquotedblright wobbling\textquotedblright%
\ mode and its hydrodynamic dissipation\cite{NOTE}.

For very short MTs we should expect deviation from this simple interpretation.
In this regime the linearly scaling internal dissipation,coming from the
migration of polymorphic defects, should start to dominate over pure
hydrodynamic friction and for sufficiently short MT lengths $L\rightarrow0$ we
could measure $\xi_{int}$ from the limit value of $\tau_{th}/L$. It appears
that for the presently available data $L>2$ $\mu m$ \cite{Taute} this
plateau-regime is not yet fully developed, enabling us to provide only an
upper numerical estimate for the inner dissipation $\xi_{int}\lesssim
4\times10^{-17}Ns$. \begin{figure}[ptbh]
\begin{center}
\includegraphics[
height=3.0684in, width=3.2292in ]{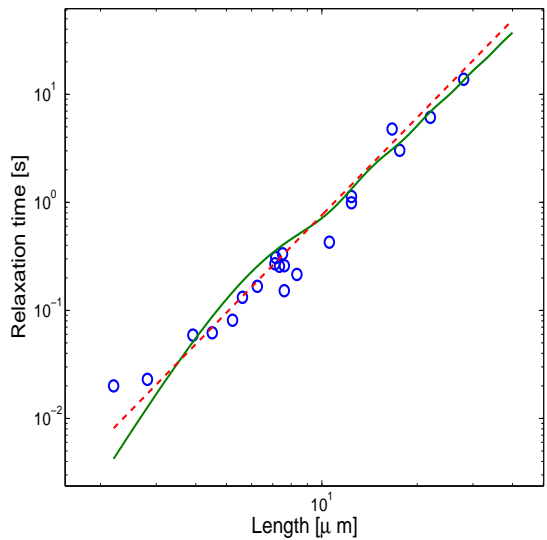}
\end{center}
\caption{The experimental microtubule relaxation times \cite{Taute} and the
no-adjustable-parameter theoretical prediction (full line) as obtained from
static data in Fig.11 (with $l_{B}=25mm,$ $\lambda=7.5\mu m,$ $\kappa_{0}%
^{-1}=18\mu m,$ $q_{0}l_{\phi}\gg1$ ). The dashed line illustrates the long
length approximation Eq. \ref{Tau3} displaying the characteristic cubic
scaling with length.}%
\end{figure}

\subsection{Very Short MTs}

Comparison with experiments for even shorter chains ($L<\pi/q_{0}$) is more
difficult due to the lack of data in clamped MT experiments ($L>2\mu m$). We
can nevertheless try a comparison with the results by the Dekker group for the
kinesin motor gliding assay of short MTs \cite{GlidingAssay,GlidingAssay2}.
Besides some similarities with the clamped fluctuating MTs, there are a number
of differences between the modelled situation of free 3D MTs and the 2D
gliding assay. In particular the 2D geometry will strongly perturb the
preferentially 3-dimensional helical ground state. The active contribution of
strong motor forces on the trajectory of short MTs is an additional potential
perturbation. Effects of MT buckling and axial MT rotation by kinesins become
likely important. This said and ignoring the differences we can still compare
(in order of magnitude) our and Dekker groups' results
\cite{GlidingAssay,GlidingAssay2}. For micron sized MTs we obtain $l_{p}%
\sim0.8\mu m$ in approximate agreement with $\sim0.2\mu m$ obtained in
\cite{GlidingAssay}. One should mention that the two cited studies
\cite{GlidingAssay,GlidingAssay2} give strongly different results depending on
whether free gliding or gliding assay with additional electric field were
considered. This difference comes from the larger deformations induced by the
electric field. Therefore the comparison of our theory with the passive
gliding assay appears more appropriate and gives a closer agreement.

\section{Summary}

In summary, we have suggested a new model that connects some of the most
persistent and confusing experimental findings concerning microtubules.
Starting from a rather broad spectrum of (apparently) disconnected
observations we have progressively built the case for a new hypothesis : the
existence of an internal switching of the GDP tubulin dimer within the
microtubule lattice. Why do microtubules become helically wavy, why do they
switch to permanently bent circular states, why do they fluctuate anomalously
when clamped? These three dangling questions became the central pillars for
the present model. Surprisingly, the simple assumption of a bistable
GDP-tubulin seems to explain these otherwise disparate phenomena in a unified
manner. As we know from recent experiments - the bistability hypothesis of
taxol stabilized protofilaments is an empirical fact indeed \cite{Multistable
Tub EM}. We have shown here, that the incorporation of such a bistable tubulin
into a closed elastic lattice changes its free behavior - it introduces strong
conformational competition among the tubulin dimers. Tubulin units on opposite
sides of the tube now start to compete for who is going to switch to the
curved state. The lattice induced frustration does not allow all the tubulin
dimers to minimize their energy individually and to switch to their preferred
states at the same time. The symmetry breaking induced by this frustration
mechanism leads to a global microtubule lattice curving. Remarkably the
curving direction is chosen randomly - and this has profound consequences. The
microtubule can chose between many energetic ground states (as many states as
protofilaments in the lattice). When we graft one end of the microtubule onto
a substrate while still allowing it to chose its bending direction freely the
strange energy degeneracy generates a very unusual thermal motion. In this
case the microtubule's motion follows -roughly speaking- a cone and it rotates
- or "wobbles"- at no energy cost around its attachment tangent. This mode of
motion - which is not to be confused with material frame rotation (which is
strictly prohibited by grafting) - is probably among the most striking
outcomes of the two state GDP-tubulin model. It is exactly this behavior that
allows to consistently explain the measurements of unusual lateral
fluctuations of grafted microtubules \cite{Pampaloni,Taute}.

\section{Perspectives}

We have focussed here on modelling taxol stabilized microtubules and the
question naturally arises if the model proposed here affects the 'real' in
vivo microtubules. On the one hand the 'weakly curved' state that is involved
in the soft polymorphic dynamics as described here seems to be (so far) a
specific signature of taxol stabilized GDP tubulin state. On the other hand we
have argued that the naturally occurring 'high curvature' GDP-tubulin state
could coexist with the straight state in the lattice under in vivo conditions
where MTs are stabilized with MAPs. The involvement of this 'high curvature'
state switching seems to manifest itself in motor driven straight to wavy
transitions of MTs in many living cell systems
\cite{Brangwyne,BicekInvivo,BrangwynneWavyBucking,Borisy,Kaech,SamsonovTau}. A
particularly impressive instance of such a polymorphic switching event in vivo
could be found in the process of axonal retraction where the whole MT
cytoskeleton of the axon undergoes a straight to helical transition and in
turn retracts towards the soma \cite{Baas}. To understand these dramatic
transitions in vivo the present theory has to be advanced and modified in two
manners. First, the effect of large active motor forces (rather than thermal
ones) has to be taken into account. In particular, one expects that under
strong buckling forces even thermodynamically unfavorable states can become
activated and constitute the ground state upon large loads.

Second, in virtually all in vivo experiments MTs are essentially confined in
2D as the containing cells adsorb to the glass substrate and assume a vary
flat 'fried egg' configuration. Consequently, the measured properties will not
necessarily reflect the three-dimensional properties of the molecule. This is
particularly important for a MT transformed to a polymorphic helix state where
the confinement entails naturally a strong deformation (of the initially 3
dimensional ground state). Under confinement the helical bending and torsional
modes become strongly coupled and bring about new physical effects. In
particular a torsionally very soft helix will have a tendency to unwind and
form in extreme cases circular arcs - reminiscent of the rings observed in
gliding assay experiments \cite{Amos}.

Finally, the local action of molecular motors could trigger a switching to the
highly curved state. While for classical motors like kinesin 1 direct evidence
for such a mode of action is still missing its relative kinesin 13
\cite{MCAK,Multistable Tub EM} has a well documented ability to actively
trigger radial bending of protofilaments. Other molecules like katanin
\cite{Katanin,Katanin2} have also been suggested to perturb the lattice and
trigger longer range transitions \cite{KulicMTshear}. This opens the
intriguing question : could classical motors (kinesin 1 and dynein) trigger
cooperative state transitions and even transmit conformational signals along
the tube? Considering the present model for stabilized MTs (where high
cooperativity is inherent to the data interpretation) this idea might not be
far fetched. In fact some evidence towards long range cooperativity of kinesin
binding along the MT was presented by Muto et al.
\cite{KinesinCooperativeBinding} - however these results still await robust reproduction.

This brings us to the question of what experiments should be performed in
order to nail down the polymorphic mechanism or any other mechanism for MT
dynamics. With microtubules being such delicate, subtle and possibly long
range correlated objects (as suggested here) a general rule of thumb for
experiments should be: Treat them more gently (do not confine) and observe
more carefully (look for correlated motion). A simple yet important experiment
would be a systematic direct observation of one-side grafted but otherwise
\emph{completely unconfined} MTs fluorescently labelled \emph{along their full
contour length}. As mentioned the presence of a quasi 2D confinement in thin
chambers as used in most MT experiments so far would perturb the native
helical MT state and should be therefore explicitly avoided. The freely
suspended gold-EM nanogrid attachment geometry as used by Pampaloni et al.
seems particularly suited for that task. Going beyond Pampaloni et al. who
labelled and traced the MT end only (via a bead), the microtubules should be
visualized along the full contour in this geometry. Tracing of several or all
points along the contour should reveal the predicted sinusoid- helical nature
of MT states. The present model predicts a peculiar cooperatively rearranging
helix state with characteristic tell tale curvature correlations between
different lattice positions which are entirely absent for usual semiflexible
filaments. Directly observing such collective motions -like the suggested
"wobbling mode" - while prohibiting trivial spacial rotations that could mask
the effects (by MT grafting) would constitute smoking gun evidence for a
polymorphism related mechanism.

In conclusion, we have proposed a novel model for internal MT lattice
dynamics. We have shown that it accounts for the otherwise mysterious MT
helicity \cite{Venier}, the anomalous length dependent lateral fluctuation
static \cite{Pampaloni,Taute} and dynamic scaling\ \cite{Taute}. The latter
two phenomena appear as mere consequences of the peculiar \textquotedblright
wobbling motion\textquotedblright\ of the polymorphic cooperatively switching
MT lattice. Although most of the observations discussed here are made in vitro
on taxol-stabilized MTs, we provided arguments in favor the existence of
polymorphic MT states in vivo. We speculate that the implied conformational
bistability of tubulin and the allosteric interaction are more than just
nature's way to modulate the elastic properties of its most important
cytoskeletal mechano-element. It could also be a missing piece in the puzzle
of polymerization "catastrophes". Even more intriguingly the predicted
structural cooperativity could allow for long range conformational signalling
along single MTs and turn the latter into an efficient "confotronic" wire
transmitting regulatory signals across the cell.

\section{Acknowledgements}

We acknowledge fruitful discussions with Francesco Pampaloni, Denis
Chr\'{e}tien, Thomas Surrey, Francois N\'{e}d\'{e}lec, Jean-Francois Joanny,
Sergey Obukhov, Linda Amos, Andr\'{e} E.X. Brown and thank Falko Ziebert for
discussion and useful comments on the manuscript.

\section{Appendix}

\subsection{A. Polymorphic phase coherence length}

In this section, we derive the formula $l_{\phi}=\frac{N^{2}b}{8\pi^{2}%
}\left(  2+e^{2J/k_{B}T}\right)  $ for the polymorphic phase coherence length.
To this end we want to calculate the distribution of double junctions that
leads to angular orientation change $\Delta\Phi$ on a scales $l$ much larger
than the tubulin dimer $b$, yet still much smaller than the total length:
$b\ll l\ll L.$ In this domain, at each cross section we have 3 possibilities:

1) State $j=0$ with no double defect. The rotation angle $\Delta\Phi$ is
attached to the internal lattice rotation,\ $\frac{\Delta\Phi}{b}-q_{0}=0$

2)\ State $j=-1$ for a left handed double defect, $\Delta\Phi$ deviates from
the internal twist :\ $\frac{\Delta\Phi}{b}-q_{0}=-\frac{1}{b}\frac{2\pi}{N}$

3)\ State $j=+1$ for a right handed double defect with $\frac{\Delta\Phi}%
{b}-q_{0}=+\frac{1}{b}\frac{2\pi}{N}.$

On a length $l$ we are throwing a 3 sided dice $l/b$ times and the total
rotation of $\Delta\Phi$ away from optimal twist is $\Delta\Phi-q_{0}%
l=\frac{2\pi}{N}\sum_{n=1}^{l/b}j_{n}.\ $The variation of the polymorphic
phase with respect to the internal twist is then%
\[
\Delta\phi=\frac{\Delta\Phi}{l}-q_{0}=\frac{1}{l}\frac{2\pi}{N}\sum
_{n=1}^{l/b}j_{n}%
\]
For $l/a>>1$ the law of large numbers implies that the random variable
$\Delta\phi=\frac{1}{l}\frac{2\pi}{N}\sum_{n=1}^{l/b}j_{n}$ becomes Gaussian
distributed
\[
p\left(  \Delta\phi\right)  \propto e^{-\frac{\Delta\phi^{2}}{2\left\langle
\Delta\phi^{2}\right\rangle }}%
\]
with mean $\left\langle \Delta\phi\right\rangle =0$ and $\left\langle
\Delta\phi^{2}\right\rangle =\left(  \frac{1}{l}\frac{2\pi}{N}\right)
^{2}\left(  \frac{l}{b}\right)  \left\langle j^{2}\right\rangle $ (as
$\left\langle j_{n}j_{m}\right\rangle =\delta_{nm}\left\langle j^{2}%
\right\rangle $). The average $\left\langle j^{2}\right\rangle $ is given from
the Boltzmann factors of the three different states $p_{0}=\frac
{1}{1+2e^{-2\beta J}}$ and $p_{\pm1}=\frac{e^{-2\beta J}}{1+2e^{-2\beta J}}$
so that $\left\langle j^{2}\right\rangle =\frac{2e^{-2\beta J}}{1+2e^{-2\beta
J}}$. We can now interpret the quantity $1/\left(  2\left\langle \Delta
\phi^{2}\right\rangle \right)  $ as coming from an effective elastic energy
over the interval $l$ by writing $\frac{\Delta\phi^{2}}{2\left\langle
\Delta\phi^{2}\right\rangle }=\frac{1}{2}\beta C_{\phi}l\left(  \Delta
\phi\right)  ^{2}$ which allows us to identify the effective stiffness
\[
C_{\phi}=kT\frac{N^{2}}{8\pi^{2}}\left(  2+e^{2J\beta}\right)  b.
\]
Note that this expression is valid for large enough $J\ $suppressing higher
order defects i.e. in the limit when multiple double defects sitting on a
single lattice site (i.e. $\left\vert j\right\vert >1$)\ can be ignored.

\subsection{B. The variation of the polymorphic modulus}

In this appendix we compute the energy variation due to a deviation of the
polymorphic modulus $\left\vert P\right\vert $ away from its optimal value
$\left\vert P^{\ast}\right\vert $ minimizing the energy; i.e., the change of
the number of switched PFs. We start with the energy density of a MT cross section%

\begin{equation}
e=\frac{B}{2}\left(  \left(  \kappa-\kappa_{pol}(p)\right)  ^{2}+\kappa
_{1}^{2}\left(  \gamma\frac{\pi}{N}p-\sin^{2}\left(  \frac{\pi}{N}p\right)
\right)  \right)
\end{equation}
whose minimum energy is reached for $p^{\ast}=\frac{N}{2}-\frac{N}{2\pi
}\arcsin\gamma$. Assuming a continuous number of PFs, the energy of a state
with $p=p^{\ast}+\Delta p$ switched PFs reads to quadratic order:%

\begin{align*}
e(p^{\ast}+\Delta p)  &  \approx e(p^{\ast})-\frac{\pi^{2}B}{N^{2}}\kappa
_{1}^{2}\cos\left(  \frac{2\pi}{N}p^{\ast}\right)  \Delta p^{2}\\
&  =e(p^{\ast})+\frac{\pi^{2}B}{N^{2}}\kappa_{1}^{2}\sqrt{1-\gamma^{2}}\Delta
p^{2}%
\end{align*}
where we used $\cos(\pi-\arcsin\gamma)=-\sqrt{1-\gamma^{2}}$. Therefore the
energy variation of a segment of length $l$ reads
\[
\Delta E\approx\frac{\pi^{2}B}{N^{2}}\kappa_{1}^{2}\sqrt{1-\gamma^{2}}%
{\textstyle\int_{0}^{l}}
ds(p(s)-p^{\ast}(s))^{2}%
\]
Now using $\left\vert P\left(  s\right)  \right\vert =\left\vert \sin\left(
\frac{\pi}{N}p\right)  \right\vert /\sin(\pi/N)$ we can write the energy
variation to the same (quadratic) order as
\[
\Delta E\approx B\kappa_{1}^{2}\sin^{2}(\pi/N)\sqrt{1-\gamma^{2}}%
{\textstyle\int_{0}^{l}}
ds((\left\vert P\left(  s\right)  \right\vert -\left\vert P^{\ast}\right\vert
))^{2}%
\]
Therefore any deviation of $P$ from its optimum state $P^{\ast}$ is associated
with an energy cost proportional to the length $l$ of the region in the
unfavorable state.

\subsection{C. Persistence length(s)}

A definition of the persistence length, often used in single molecule
experiments, is expressed in terms of the lateral deviation $\overrightarrow
{\rho}=(x(s),y\left(  s\right)  )$ of a MT clamped at $s=0$ from its
attachment axis : $l_{p}^{\ast}\left(  s\right)  =2/3s^{3}/\left\langle
\left(  \overrightarrow{\rho}\left(  s\right)  -\left\langle \overrightarrow
{\rho}\left(  s\right)  \right\rangle \right)  ^{2}\right\rangle $ and
$\left\langle ..\right\rangle $ is the statistical average. The equivalence of
the $x$ and $y$ directions implies that $l_{p}^{\ast}\left(  s\right)
=1/3s^{3}/\left\langle \left(  y\left(  s\right)  -\left\langle y\left(
s\right)  \right\rangle \right)  ^{2}\right\rangle .$ The second often used
alternative but more standard definition of the persistence length - the
tangent persistence length- is related to the angular correlation
$l_{p}\left(  s-s^{\prime}\right)  =\left\vert s-s^{\prime}\right\vert
/V(s-s^{\prime})$ with the variance $V=\left\langle \left(  \theta_{y}\left(
s\right)  -\theta_{y}\left(  s^{\prime}\right)  \right)  ^{2}\right\rangle $
(by symmetry we have the same expression with $\theta_{x}$). Whereas for an
ideal WLC $l_{p}^{\ast}=l_{p}=l_{B}$ is position and definition independent
this is not the case for a polymorphic chain (see Fig. 13). For small angular
deformations the decoupling of chain's fluctuations into polymorphic and
purely elastic contributions allows to decompose the persistence length as
$l_{p}^{-1}=l_{pol}^{-1}+l_{B}^{-1}$ - this result being valid for both
definitions of the persistence length.

We first focus on the first definition - the clamped persistence length. In
this case the polymorphic persistence length $l_{pol}^{\ast}(s)$ is given by%

\begin{equation}
l_{pol}^{\ast}(s)=1/3s^{3}/\left\langle \left(  y_{pol}\left(  s\right)
-\left\langle y_{pol}\left(  s\right)  \right\rangle \right)  ^{2}%
\right\rangle \label{lpol1}%
\end{equation}
where $y_{pol}(s)$ is the lateral polymorphic displacement in the $y$
direction. Integrating over the rotational zero mode readily implies
$\left\langle y_{pol}\left(  s\right)  \right\rangle =0$ (see Eq. \ref{lpol}).
From Eq. \ref{ypolsquare} one can write%

\[
\left\langle y_{pol}^{2}\left(  s\right)  \right\rangle =\int_{0}^{s}\int
_{0}^{s}G(s_{1},s_{2})ds_{1}ds_{2}%
\]
with the angular correlation function $G(s_{1},s_{2})=\left\langle
\theta_{y,pol}(s_{1})\theta_{y,pol}(s_{2})\right\rangle $ given by the
integration over the zero mode
\begin{equation}
G(s_{1},s_{2})=\int_{0}^{2\pi}\frac{d\phi_{0}}{2\pi}G_{0}(s_{1},s_{2},\phi
_{0}) \label{G}%
\end{equation}
of the angular correlation function at fixed value of $\phi_{0}$, i.e.,
$G_{0}(s_{1},s_{2},\phi_{0})=\left\langle \theta_{y,pol}(s_{1})\theta
_{y,pol}(s_{2})\right\rangle |_{\phi_{0}}$. This last expression, from the
relation $\theta_{y,pol}(s)=\kappa_{0}\int_{0}^{s}\sin\left(  \widetilde{\phi
}\left(  s^{\prime}\right)  +q_{0}s^{\prime}+\phi_{0}\right)  ds^{\prime}$
(cf. Eq. \ref{Aphi}) is explicitly given by
\begin{align}
G_{0}(s_{1},s_{2},\phi_{0})  &  =\kappa_{0}^{2}\int_{0}^{s_{1}}\int_{0}%
^{s_{2}}\left\langle \sin\left(  \widetilde{\phi}\left(  s\right)
+q_{0}s+\phi_{0}\right)  \right. \nonumber\\
&  \left.  \sin\left(  \widetilde{\phi}\left(  s^{\prime}\right)
+q_{0}s^{\prime}+\phi_{0}\right)  \right\rangle |_{\phi_{0}}dsds^{\prime}.
\end{align}
After integration over $\phi_{0}$ and using the known result $\left\langle
\cos\left(  \widetilde{\phi}\left(  s_{1}\right)  -\widetilde{\phi}\left(
s_{2}\right)  \right)  \right\rangle =e^{-\left\vert s_{1}-s_{2}\right\vert
/2l_{\phi}}$ which results from the WLC type probability distribution of the
field $\widetilde{\phi}$, i.e., $P[\widetilde{\phi}]\sim\exp(-\frac{l_{\phi}%
}{2}\int_{0}^{L}ds\widetilde{\phi}^{\prime2})$ one obtains the rotational
invariant correlation function in the form%

\begin{equation}
G(s_{1},s_{2})=\frac{\kappa_{0}^{2}}{2}\int_{0}^{s_{1}}\int_{0}^{s_{2}%
}e^{-\frac{\left\vert s-s^{\prime}\right\vert }{2l_{\phi}}}\cos\left(
q_{0}(s-s^{\prime})\right)  dsds^{\prime} \label{angular}%
\end{equation}
Computation of the integrals in Eq. \ref{angular} gives finally the following
expression for the polymorphic contribution of the transverse displacement
\begin{align}
\left\langle y_{pol}(s)^{2}\right\rangle  &  =\frac{2\kappa_{0}^{2}l_{\phi}%
}{3(1+4l_{\phi}^{2}q_{0}^{2})^{4}}\left\{  P_{1}-e^{-\frac{s}{2l_{\phi}}}%
P_{2}\cos\left(  q_{0}s\right)  \right. \nonumber\\
&  \left.  +e^{-\frac{s}{2l_{\phi}}}P_{3}\sin(q_{0}s)\right\}  \label{lateral}%
\end{align}
with ${\small P}_{1}(s){\small =24l}_{\phi}^{3}\left(  1-6x+x^{2}\right)
{\small -3l}_{\phi}\left(  1+x-x^{2}-x^{3}\right)  {\small s}^{2}$

${\small +}\left(  1+3x+3x^{2}+x^{3}\right)  {\small s}^{3},$ ${\small P}%
_{2}(s){\small =24l}_{\phi}^{3}\left(  1-6x+x^{2}\right)  $

${\small +12l}_{\phi}^{2}\left(  1-2x-3x^{2}\right)  {\small s}$ and
${\small P}_{3}(s){\small =192l}_{\phi}^{4}{\small q}_{0}{\small (1-x)}$

${\small +24l}_{\phi}^{3}{\small q}_{0}{\small (3+2x-x}^{2}{\small )s}$ where
we have introduced the notation $x=4l_{\phi}^{2}q_{0}^{2}.$

From Eq. \ref{lateral}, we get the polymorphic persistence length
$l_{pol}^{\ast}(s)$ defined in Eq. \ref{lpol1}, and in turn the global
persistence length $l_{p}^{\ast}(s)$ depicted in Fig. 13. Its physical
interpretation is discussed in the main text.

We now consider the second definition of the persistence length $l_{p}\left(
s-s^{\prime}\right)  =\left\vert s-s^{\prime}\right\vert /V(s-s^{\prime})$.
From Eq. \ref{angular}, the angular variance $V_{pol}$ can easily be
evaluated
\begin{align}
V_{pol}(s)  &  =\frac{2\kappa_{0}^{2}l_{\phi}}{1+4l_{\phi}^{2}q_{0}^{2}%
}\left(  s-\frac{2l_{\phi}\left(  1-4q_{0}^{2}l_{\phi}^{2}\right)
}{1+4l_{\phi}^{2}q_{0}^{2}}\right) \nonumber\\
&  +\frac{4\kappa_{0}^{2}l_{\phi}^{2}e^{-\frac{s}{2l_{\phi}}}}{\left(
1+4l_{\phi}^{2}q_{0}^{2}\right)  ^{2}}\left(  \left(  1-4q_{0}^{2}l_{\phi}%
^{2}\right)  \cos\left(  q_{0}s\right)  \right. \nonumber\\
&  \left.  -4q_{0}l_{\phi}\sin\left(  q_{0}s\right)  \right)  \label{C3}%
\end{align}
The resulting persistence length $l_{p}$ (depicted in Fig. 13) shows a rich
behavior similar to the persistence length $l_{p}^{\ast}\left(  s\right)  $
but displays a distinct functional form from the latter. However as expected,
both curves reach the same asymptotic value at very short and very long MT lengths.

\begin{figure}[ptb]
\begin{center}
\includegraphics[
height=2.6in]{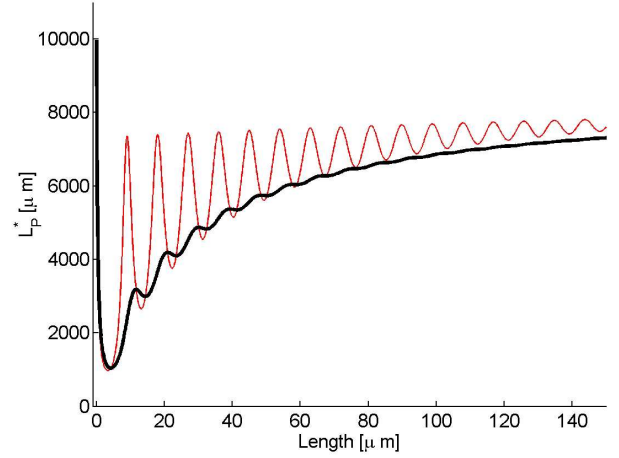}
\end{center}
\caption{Different definitions of the persistence length can deviate from each
other for a polymorphic chain. The "clamped persistence length" $l_{p}^{\ast}$
(thick line) and the "tangent persistence length" $l_{p}$ (thin line) (for
$l_{B}=10mm,$ $l_{\phi}=50\mu m,$ $\kappa_{0}=0.03\mu m^{-1}$ and
$q_{0}=0.7\mu m^{-1}$).}%
\end{figure}

\subsection{D. Zero mode dynamics}

The evolution of the zero mode $\phi_{0}\left(  t\right)  $ is given by Eq.
\ref{phiotime}:%

\begin{equation}
\frac{d}{dt}\phi_{0}\left(  t\right)  =\frac{1}{\xi_{tot}}L^{-1}\int_{0}%
^{L}\Gamma_{\phi}\left(  s,t\right)  ds \label{dphiodt}%
\end{equation}
with a friction constant $\xi_{tot}=\xi_{int}+\xi_{ext}$ where $\xi_{ext}$ is
given by Eq. \ref{Xiexternal}. The correlation function of the thermal white
$\Gamma_{\phi}\left(  s,t\right)  $ noise is $\left\langle \Gamma_{\phi
}\left(  s,t\right)  \Gamma_{\phi}\left(  s^{\prime},t^{\prime}\right)
\right\rangle =D\delta(s-s^{\prime})\delta(t-t^{\prime})$ with a coefficient
$D$ that can be determined in the following manner. Notice first that
$\phi_{0}$ performs a \ free Brownian motion and its quadratic fluctuations
necessarily satisfy the relation $\left\langle \left(  \phi_{0}\left(
t\right)  -\phi_{0}\left(  0\right)  \right)  ^{2}\right\rangle =\frac
{2k_{B}T}{L\xi_{tot}}t$. On another hand integrating Eq. \ref{dphiodt} :%
\begin{equation}
\phi_{0}\left(  t\right)  -\phi_{0}\left(  0\right)  =\xi_{tot}^{-1}%
{\displaystyle\int\nolimits_{0}^{t}}
\left(  \frac{1}{L}\int_{0}^{L}\Gamma\left(  s,t^{\prime}\right)  ds\right)
dt^{\prime} \label{phiT2}%
\end{equation}
and averaging over the white noise the quadratic phase fluctuations one
obtains : $\left\langle \left(  \phi_{0}\left(  t\right)  -\phi_{0}\left(
0\right)  \right)  ^{2}\right\rangle =\frac{D}{\xi_{tot}^{2}L}t,$ from which
we readily deduce $D=2\xi_{tot}k_{B}T$ - as expected from the fluctuation
dissipation theorem.

The relaxation time is generally given from the time correlation function
$<y_{pol}(s,0)y_{pol}(s,t)>$\ with the lateral position $y_{pol}%
(s,t)=\frac{\kappa_{0}}{q_{0}^{2}}(sq_{0}\cos\left(  \phi_{0}\left(  t\right)
+\alpha\right)  +\sin\left(  \phi_{0}\left(  t\right)  +\alpha\right)
-\sin\left(  q_{0}s+\phi_{0}\left(  t\right)  +\alpha\right)  )$\ obtained
from Eq. \ref{Polydeplacement} with $l_{\phi}>>s$. The average must first take
into account all statistically equivalent values of angular orientations
$\alpha\in\left[  0,2\pi\right]  ,$\ such that $\left\langle y_{pol}%
(s,0)y_{pol}(s,t)\right\rangle =\int\nolimits_{0}^{2\pi}\left\langle
y_{pol}(s,0)y_{pol}(s,t)\right\rangle _{\alpha}\frac{d\alpha}{2\pi}$\ and we
obtain
\begin{equation}
\left\langle y_{pol}(s,0)y_{pol}(s,t)\right\rangle =\left\langle y_{pol}%
^{2}(s)\right\rangle \left\langle \cos(\phi_{0}\left(  t\right)  -\phi
_{0}\left(  0\right)  )\right\rangle
\end{equation}
with $<y_{pol}^{2}(s)>=\frac{\kappa_{0}^{2}}{q_{0}^{2}}\left(  \frac{s^{2}}%
{2}+\frac{1-\cos(q_{0}s)}{q_{0}^{2}}-\frac{s\sin(q_{0}s)}{q_{0}}\right)
$\ corresponding to the static result Eq. \ref{lateral} in the limit
$s/l_{\phi}<<1$. With $\ref{phiT2}$\ defining a simple Gaussian random walk
processes one straightforwardly obtains%

\begin{equation}
\left\langle \cos(\phi_{0}\left(  t\right)  -\phi_{0}\left(  0\right)
)\right\rangle =e^{-t/\tau}%
\end{equation}
with the relaxation time given by
\begin{equation}
\tau=L\frac{\xi_{tot}}{k_{B}T}.
\end{equation}

\subsection{E. Comment on MT surface attachment and the robustness of
"wobbling"}

Throughout this work we have assumed that the free rearrangement of the
polymorphic lattice states is not significantly hindered by the covalent
surface attachment of the MT, as e.g. performed in \cite{Pampaloni} and
\cite{Taute}. This assumption is integral to the "wobbling" motion and in turn
to understanding the static and dynamic data scaling. It therefore deserves a
closer consideration.

In the experiments by Pampaloni \cite{Pampaloni} et al. and Taute et al.
\cite{Taute} the adsorbed MT part is attached to a gold (electron microscopy
grid) surface via thiol groups. It is likely that $\approx$ 1 - 2
protofilaments will establish localized chemical contacts with the gold
microgrid. While a substantial perturbation of the dimer like e.g.
denaturation appears unlikely, it is unclear to what extent this procedure
will perturb the inner (polymorphic) dynamics of the entire tubulin dimer
units. In principle one can anticipate two plausible scenarios that would to a
varying degree interfere with the free "wobbling" motion:

S1) Due to high cooperativity (large coupling $J$) the polymorphic state
transition can propagate within a certain penetration depth into the adsorbed
(straight-planar) MT section.

S2) The cooperativity is too weak to compete with the constraints imposed by
the surface (including chemical perturbations) and the polymorphic transition
does not propagate into the straight adsorbed MT section.

In both cases we have a non vanishing deflection angle between a forced
(adsorbed) planar section and the free helical section direction - causing
effectively the characteristic MT "kink" at the surface interface. However the
rotational mobility of this "kink" (wobbling mode) which is integral to our
theory will be affected in slightly different manner.

If in case S1 in the adsorbed section the polymorphic order parameter P can
rearrange to some extent (by switching the monomer states without causing a
detectable deformation) except for possibly in the few surface interacting
dimers, then the effects of the "wobbling" motion will be hindered only mildly
in the following sense. To retrieve the anomalous lateral fluctuations it is
indeed enough for the wobbling angle $\phi_{0}$ to move freely in a certain
non-vanishing angular interval. A single complete or multiple rotations of the
order parameter $\vec{P}$ are not strictly necessary for the "hinge" effect -
and they are in fact equivalent in lateral projection (as in experiment) to
the motion of the wobbling angle $\phi_{0}$ in the smaller interval
$[-\pi/2,+\pi/2]$. Note that even smaller intervals than that will lead to a
similar phenomenology (in particular dynamic and static variable scalings with
length). Thus the conical hinge-like motion is in a sense robust with respect
to a limited local rotational hindrance perturbation in the adsorbed region.

In the scenario S2 the situation is somehow simpler as the polymorphic
dynamics of the adsorbed region is not involved in the process (the
polymorphic order parameter vanishes there: $\vec{P}=0$). Wobbling is realized
through a coherent rearrangement of the free MT section alone - without a
strong coupling to the adsorbed region.

Although both attachment scenarios S1 and S2 appear to some extent plausible,
at present it is difficult to make reliable statements about their respective
likelihood. In fact only a posteriori we can cautiously state that based on
the experimental static and dynamic measurement evidence: the chain "wobbles"
to a high enough extent to display the effects that we observe.


\begin{thebibliography}{99}
\bibitem {NOTE}In Appendix E we discuss possible effects induced by surface
attachment that could to some extent interfere with the ideal free "wobbling"
motion in experiments.
\end{thebibliography}
\end{document}